\documentclass{cta-author}

\usepackage{amsmath}
\usepackage{amssymb}
\usepackage{bm}
\usepackage{multirow}
\usepackage{color}
\usepackage{graphicx}
\usepackage[ruled,commentsnumbered]{algorithm2e}
\usepackage{algorithmic}
\usepackage{url}
\usepackage{acro}

\newlength\myindent
\newcommand\blfootnote[1]{%
  \begingroup
  \renewcommand\thefootnote{}\footnote{#1}%
  \addtocounter{footnote}{-1}%
  \endgroup
}

\def\w{{\bm{w}}}
\def\g{{\bm{g}}}
\def\x{{\bm{x}}}
\def\s{{\bm{s}}}
\def\y{{\bm{y}}}
\def\vv{{\bm{v}}}
\def\yH{{\bm{y}}^H}
\def\wH{{\bm{w}}^H}
\def\C{{\bm{C}}}
\def\B{{\bm{B}}}
\def\hy{\widehat{\bm{y}}}

\DeclareMathOperator*{\argmin}{arg\,min}
\DeclareMathOperator*{\argmax}{arg\,max}

\DeclareAcronym{DNN}{
	short = DNN ,
	long  = \acspace\acspace Deep Neural Network ,
	class = abbrev
}
\DeclareAcronym{BSS}{
	short = BSS ,
	long  = \acspace\acspace Blind Source Separation ,
	class = abbrev
}
\DeclareAcronym{DSB}{
	short = DSB ,
	long  = \acspace\acspace Delay-And-Sum Beamformer ,
	class = abbrev
}
\DeclareAcronym{SDW-MWF}{
	short = SDW-MWF ,
	long  = \acspace\acspace Speech Distortion Weighted Multi-Channel Wiener Filter ,
	class = abbrev
}
\DeclareAcronym{MMSE}{
	short = MMSE ,
	long  = \acspace\acspace Minimum Mean Square Error beamformer ,
	class = abbrev
}
\DeclareAcronym{MVDR}{
	short = MVDR ,
	long  = \acspace\acspace Minimum Variance Distortionless Response ,
	class = abbrev
}
\DeclareAcronym{MSNR}{
	short = MSNR ,
	long  = \acspace\acspace Maximum Signal-to-Noise-Ratio beamformer ,
	class = abbrev
}
\DeclareAcronym{GEV}{
	short = GEV ,
	long  = \acspace\acspace Generalised EigenValue beamformer ,
	class = abbrev
}
\DeclareAcronym{RTF}{
	short = RTF ,
	long  = \acspace\acspace Relative Transfer Function ,
	class = abbrev
}
\DeclareAcronym{VAD}{
	short = VAD ,
	long  = \acspace\acspace Voice Activity Detector ,
	class = abbrev
}
\DeclareAcronym{ICA}{
	short = ICA ,
	long  = \acspace\acspace Independent Component Analysis ,
	class = abbrev
}
\DeclareAcronym{WER}{
	short = WER ,
	long  = \acspace\acspace Word Error Rate ,
	class = abbrev
}
\DeclareAcronym{IRTF}{
	short = IRTF ,
	long  = \acspace\acspace Inverse Relative Transfer Function beamformer ,
	class = abbrev
}
\DeclareAcronym{BAN}{
	short = BAN ,
	long  = \acspace\acspace Blind Analytic Normalisation post-filter ,
	class = abbrev
}
\DeclareAcronym{STFT}{
	short = STFT ,
	long  = \acspace\acspace Short-Time Fourier Transform ,
	class = abbrev
}
\DeclareAcronym{BLSTM}{
	short = BLSTM ,
	long  = \acspace\acspace Bi-Directional Long Short-Term Memory ,
	class = abbrev
}
\DeclareAcronym{IBM}{
	short = IBM ,
	long  = \acspace\acspace Ideal Binary Masks ,
	class = abbrev
}
\DeclareAcronym{PSD}{
	short = PSD ,
	long  = \acspace\acspace Power Spectral Density ,
	class = abbrev
}
\DeclareAcronym{PESQ}{
	short = PESQ ,
	long  = \acspace\acspace Perceptual Evaluation of Speech Quality ,
	class = abbrev
}
\DeclareAcronym{SIR}{
	short = SIR ,
	long  = \acspace\acspace Signal-to-Interference ratio ,
	class = abbrev
}
\DeclareAcronym{SDR}{
	short = SDR ,
	long  = \acspace\acspace Signal-to-Distortion ratio ,
	class = abbrev
}
\DeclareAcronym{SAR}{
	short = SAR ,
	long  = \acspace\acspace Signal-to-Artefact ratio ,
	class = abbrev
}
\DeclareAcronym{RNNLM}{
	short = RNNLM ,
	long  = \acspace\acspace Recurrent Neural Network Language Model ,
	class = abbrev
}
\acsetup{first-style=short}

\begin{document}

\supertitle{Block-Online Multi-Channel Speech Enhancement Using DNN-Supported Relative Transfer Function Estimates}

\title{Block-Online Multi-Channel Speech Enhancement Using DNN-Supported Relative Transfer Function Estimates}

\author{\au{Jiri Malek$^{1}$}, \au{Zbyn\v{e}k~Koldovsk\'{y}$^{1}$}, \au{Marek Bohac$^{1}$}}

\address{\add{1}{Faculty of Mechatronics, Informatics, and Interdisciplinary 
Studies, Technical University of Liberec, Studentsk\'a 2, Liberec, Czech Republic}
\email{jiri.malek(at)tul.cz}}

\begin{abstract}
This work addresses the problem of block-online processing for multi-channel speech enhancement. Such processing is vital in scenarios with moving speakers and/or when very short utterances are processed, e.g., in voice assistant scenarios. We consider several variants of a system that performs beamforming supported by DNN-based voice activity detection (VAD) followed by post-filtering. The speaker is targeted through estimating relative transfer functions between microphones.
Each block of the input signals is processed independently in order to make the method applicable in highly dynamic environments. Owing to the short length of the processed block, the statistics required by the beamformer are estimated less precisely. The influence of this inaccuracy is studied and compared to the processing regime when recordings are treated as one block (batch processing).  
The experimental evaluation of the proposed method is performed on large datasets of CHiME-4 and on another dataset featuring moving target speaker. The experiments are evaluated in terms of objective and perceptual criteria (such as signal-to-interference ratio (SIR) or perceptual evaluation of speech quality (PESQ), respectively). Moreover, word error rate (WER) achieved by a baseline automatic speech recognition system is evaluated, for which the enhancement method serves as a front-end solution. 
The results indicate that the proposed method is robust with respect to short length of the processed block. Significant improvements in terms of the criteria and WER are observed even for the block length of $250$~ms. 
\end{abstract}

\maketitle

\section{Introduction}
\label{sec:intro}
\blfootnote{This paper is a post-print of a paper submitted to and accepted for publication in IET Signal Processing and is subject to Institution of Engineering and Technology Copyright. The copy of record is available at the IET Digital Library}
Speech enhancement in real-world conditions is an open audio signal processing problem with many applications ranging from distant communication and automatic speech recognition to hearing aids.
The enhancement aims at the suppression of distortions present in target speech recordings, such as background noise, reverberation or cross-talk. It is a challenging problem as real-world environments involve various situations, rooms, interferences, transient sounds and non-stationary noises. In particular, it is difficult to develop systems that can efficiently operate in general conditions with low processing delays~\cite{beambook}.

One of the basic distinctions among enhancement techniques is the number of input signals they use. Single-channel techniques rely only on spectral filtering, while multi-channel methods also exploit the spatial information.
Single channel enhancement is constituted by classical techniques such Spectral Subtraction~\cite{singlech_subtr,singlech_subtr_iet} or Minimum Mean Square Estimator~\cite{singlech_mmse}, as well as modern techniques based either on statistical principles~\cite{singlech_lowdist_mmse, singlech_newortho_supergauss} or
supported by machine learning (e.g.,~\cite{singlech_dnn} and overview in~\cite{singlech_dnn_overview}).
Due to more information available, multi-channel filters usually achieve better enhancement compared to single-channel ones~\cite{overview}. Moreover, spatial information is important to determine the target speaker, which can be done, e.g., based on localisation.
Multi-channel approaches can be based on classical array 
processors, commonly referred to as ''beamformers'', or on blind source separation (BSS). BSS techniques try to separate individual source signals that are simultaneously active and mixed together~\cite{bssbook}. Beamformers exploit known 
parameters of the signal mixture, which, however, must also be estimated when only mixed signals are observed. For example, under the free-field 
propagation model, direction of arrival and array geometry are used to select the optimum filter coefficients. In reverberant environments, acoustic transfer functions and/or signals' spatial covariance matrices must be estimated in order to compute an optimum beamformer~\cite{beambook}.

A conventional technique is the Delay-and-Sum Beamformer 
(DSB)~\cite{dsb}, which is based on the free-field assumption and performs the enhancement through summing signals from delay-compensated channels.
More advanced beamformers take into account the reverberation.
To name the most popular approaches: The Speech Distortion Weighted Multi-Channel Wiener Filter (SDW-MWF) minimises 
the square distance between filtered inputs and the unknown 
target~\cite{sdwmwf}. It involves a free parameter which provides a 
trade-off between noise suppression and speech distortion.
A particular choice of this parameter leads to the classical Minimum Mean 
Square Error (MMSE) beamformer~\cite{vantrees}, which is sometimes referred to as the Multi-channel Wiener filter~\cite{overview}. 
For another choice of the parameter, the Minimum Variance Distortionless 
Response (MVDR) beamformer is obtained~\cite{mvdrmwfbook}.
The Maximum Signal-to-Noise (MSNR) beamformer attempts to maximise the SNR in the output signal~\cite{msnr}. Its application in enhancement of speech was 
presented in~\cite{umbach_gev}, where it was referred to as Generalised 
EigenValue beamformer (GEV). For a more comprehensive overview of the beamforming techniques, see \cite{overview}.

The beamformers rely on robust estimation of their inner 
parameters, such as covariance matrices of speech/noise or relative transfer functions (RTF).
RTF is a transfer function between two channels in response to the target source, as defined in~\cite{gannot}.
Since the speech/noise covariance values are not known in noisy situations, the methods usually exploit Voice Activity Detectors to localise signal 
frames or bins in the time-frequency domain where only noise or speech is 
active. The desired statistics are then estimated using these frames/bins. 

Parametric VAD methods rely on statistical models for speech/noisy signals 
\cite{vad_stat_chang}.
Often, the drawback resides in their limited ability to model highly non-stationary noise and transient interferences~\cite{vad_stat_drawb}.
To overcome this drawback, several approaches have been proposed. One is to focus solely on harmonic properties of voiced speech instead of assuming any specific properties of diverse noise~\cite{vad_stat_speech_iet}. Other approach is to exploit machine-learning principles~\cite{vad_dnn_zhang}, especially, deep neural networks (DNN) \cite{dnn_book}. 

Concerning the RTF estimation, various approaches exist in the literature.
The covariance subtraction~\cite{cohen2004} and covariance 
whitening~\cite{rtf_covwei} methods require the noise-speech and noise 
covariance matrices.
The method from~\cite{shalvi1996} alleviates this requirement by assuming that the noise is stationary and that the noise statistics can be estimated along with the RTF. The RTF can also be estimated using BSS; an analysis of 
the techniques based on independent component analysis (ICA) can be found, e.g., in~\cite{rtf_ica_analysis,khan2015}; see also \cite{koldovsky2015,giri2016,katzberg2018} for approaches exploiting sparsity.

Multi-channel enhancement techniques can be used as preprocessors for systems performing robust automatic speech recognition (ASR) as described in overviews~\cite{asr_overview_2013,asr_overview_2018}.

\subsection{Community-Based Campaign: CHiME}

Although the theoretical properties of the enhancement techniques discussed in Section~\ref{sec:intro} are well known,
their application in real scenarios is difficult, due to the need to robustly estimate their inner parameters.
This motivates the researchers working in the fields of speech enhancement and 
robust speech recognition to organise evaluation campaigns where the most 
recent technologies are compared in practical tasks. The campaigns thus reveal the current state-of-the-art methods.
The most recognised is the CHiME Speech Separation and Recognition Challenge (CHiME).
This work focuses on the data from the CHiME-4~\cite{chime_web} challenge, i.e., 
real-world multi-channel recordings of a single speaker originating in four distinct noisy environments.

The solutions proposed for CHiME-4 are mainly focused on the robust speech 
recognition of multi-channel data, since the evaluation proceeds in terms of WER. 
Nevertheless, the front-end processing prior the feature computation is one of the key components, 
because it can improve the final WER by as much as several percent. 
Most of CHiME-4 front-end processors combine beamforming along with information 
obtained through machine learning. For example, a Gaussian-Mixture-Model-based 
VAD is utilised in~\cite{spmi_chime4} to assist the estimation of speech/noise 
covariance matrix for MVDR. For a similar purpose, the 
paper~\cite{heymann} utilises a DNN-based VAD along with MVDR and GEV 
beamformers; see also \cite{heymann_chime4}.
In \cite{iflytek_chime4}, the enhancement is performed in two passes. First, MVDR is used to obtain 
enhanced speech, which is then forwarded to an external speech recogniser. In 
the second pass, a fine-tuned beamforming is performed using a VAD that exploits 
signal segmentation provided by the speech recogniser.

An important aspect of CHiME-4 is that, although the acoustic conditions are varying across different recordings, 
the position of the speaker is rather static within single recording. Many state-of-the-art methods proposed for CHiME-4 take advantage of this fact and use the whole recording as one batch of data to estimate 
the necessary speech/noise spatial covariance matrices as precise as possible. 
Some methods perform several passes \cite{iflytek_chime4,beamformit} in order to optimise the enhancement of the given recording. This approach is not suitable for more dynamic recordings. 
For example, when the target speaker is moving, short blocks of the signals need to be processed, in order to continuously update the necessary statistics with respect to the changing position. 
The design of a technique intended for short data segments and its comparison to techniques used to solve CHiME-4 in the batch manner are the objectives of this work.

Very recently, CHiME-5 challenge~\cite{chime5_web} was released, featuring dinner party of four speakers in a domestic scenario. This dataset contains recordings with more speaker movement. However, the experiments in this work are focused on the CHiME-4 data rather than CHiME-5. The recordings of the latter dataset contain high amount of cross-talk (up to four speaker are active simultaneously), which is beyond the scope of the current research. 

\subsection{Contribution}

We propose an online multi-channel enhancement system that processes data block-by-block. 
To maximise the adaptation speed of the system, the estimation of inner statistics is {\em not} recursive, i.e., it does not utilise estimates from previous blocks.	Such processing is vital in scenarios with moving speakers and/or when very short utterances or mere keywords are processed.
The proposed system is implemented in two variants, respectively, with the approximate Minimum Mean-Squared Error beamformer \cite{koldovsky2016} and with a robust Inverse Relative Transfer Function beamformer (IRTF).

The beamformer steering is based on explicit estimation of RTFs, as compared to the recently published supervised techniques that mainly rely on the estimation of covariance matrices. For example, some methods (e.g., \cite{heymann,heymann_chime4,araki2016,review_nakatani,review_pfeifen}) estimate the steering vectors as principal components of speech covariance matrices. The methods in \cite{review_ochiai_endtoend,review_erdogan} rely on alternative MVDR computations that solely depend on the covariance matrices of noise and of noisy data.
In our experiments, the methods based on eigenvalue decomposition of covariance matrices are represented by GEV from~\cite{heymann}. Our experiments indicate that the performance of GEV deteriorates, when short processing blocks are analysed. In contrast, the proposed method, based on direct RTF estimation, yields stable performance.
The RTF estimator used in this work is endowed with a DNN-based VAD which improves the  estimation accuracy and lowers the WER of the ASR back-end. Nevertheless, the VAD inclusion is optional and may be excluded, if, e.g., computational demands due to the VAD are of concern. 
Recently, similar solution using the RTF estimation for steering the MVDR beamformer have appeared in \cite{icassp2018_rtf_beam}. However, the paper is focused on the batch processing and the RTF estimator utilised in the work cannot be used without VAD.

The system's capabilities are validated on the CHiME-4 datasets and on a multi-channel dataset featuring moving speaker. Our study is focused on how the enhancement performance depends on the length of the processing blocks. The results for CHiME-4 are evaluated in terms of WER achieved by the baseline CHiME-4 speech recognition system. Also, the speech enhancement capabilities of the systems are quantified using objective criteria reflecting the output signal quality. 

Recently, researchers started investigating the (block)-online behaviour of DNN-endowed beamformers.
The works in~\cite{icassp2018_practical_beam}, \cite{icassp2018_online_beam},  \cite{higuchi_online_recursive_2016} and \cite{icassp2019_togami_recurbeam} present block-online/frame-by-frame VAD-based MVDR beamformers, which are all based on covariance matrix operations. In contrast to the current work, the papers neither analyse the dependence of the input length on the performance, nor the perceptual quality of the output. 

This paper is organised as follows. In Section~\ref{sec:problem}, the multi-channel signal enhancement problem is formulated and basic beamforming techniques are introduced. Section~\ref{sec:method} is devoted to a detailed description of the proposed multi-channel enhancement system. In Section~\ref{sec:experiments}, the results of extensive tests and comparisons are presented. Section~\ref{sec:conclusion} provides conclusions.

\section{Problem description}
\label{sec:problem}

\subsection{The model}

Considering the target speaker as a directional source, a noisy recording can be described in the short-time Fourier domain as
\begin{equation}
\label{model}
{\x}(k,\ell) = \underbrace{{\g}(k,\ell)s(k,\ell)}_{\s(k,\ell)} + {\y}(k,\ell),
\end{equation}
where ${\x}(k,\ell)$ is the $M\times1$ vector of noisy microphone signals, $M$ is the number of microphones, 
and $s(k,\ell)$ is the target speech as observed on a reference microphone. 
Next, ${\y}(k,\ell)$ is a vector involving all noise components, which are assumed to be uncorrelated with $s(k,\ell)$; 
$k$ represents the frequency index; and $\ell$ stands for the time frame index.

The elements of ${\g}(k,\ell)$ 
contain the RTFs of all microphones with respect to the reference one \cite{gannot}.
The speaker can move during the utterances; therefore, the RTFs can vary in time. It is assumed that the changes are slow, that is, ${\g}(k,\ell)$ remains approximately constant during each block of frames.
The spatial image of the target speech as measured on the microphones is given by ${\s}(k,\ell) = {\g}(k,\ell)s(k,\ell)$.
In the following text, the indexes $k$ and $\ell$ are omitted for the sake of brevity, where no confusion arises.

\subsection{Multi-channel processing}

This section describes the basic forms of the beamforming techniques discussed in this work. 
Let the output of a beamformer be denoted as
\begin{equation}
    u(k,\ell) \triangleq \w(k,\ell)^H\x(k,\ell),
\end{equation}
where $\w$ is the steering vector of the beamformer, which takes the following forms.

\emph{The MVDR beamformer} is a popular multi-channel processor for noise suppression but it may also be used for dereverberation, as suggested in~\cite{schwartz}. It is defined as the constrained minimiser
\begin{equation}
{\w}_{\rm MVDR} = \argmin_{\w} \{\wH \C_{\y\y} \w \ {\rm s.t.}\ \wH\g = 1 \},
\end{equation}
which gives
\begin{equation}
\label{mvdr}
{\w}_{\rm MVDR}=\frac{\C_{\y\y}^{-1}{\g}}{{\g}^H\C_{\y\y}^{-1}{\g}},
\end{equation}
where $\C_{\y\y}={\rm E}[\y\yH]$ denotes the covariance matrix of noise $\y$;
${\rm E}[\cdot]$ stands for the expectation operator, and $(\cdot)^H$ denotes the conjugate transpose.

\emph{The MMSE beamformer} is defined through
\begin{equation}
{\w}_{\rm MMSE}=\argmin_{\w} {\rm E} \{|\w^H\x-s|^2\}.
\end{equation}

In fact, MMSE can be implemented as a cascade of MVDR and a single-channel Wiener Filter~\cite{overview}. The single-channel filter suppresses the residual noise $r_{\rm MVDR}=\wH_{\rm MVDR}\y$ in the beamforming output $u$. Specifically,
\begin{equation}\label{mmse}
{\w}_{\rm MMSE}={\w}_{\rm MVDR}\underbrace{
	\frac{{\rm E}[|u|^2]-{\rm E}[|r_{\rm MVDR}|^2]}{{\rm E}[|u|^2]}
}_\textrm{Wiener post-filter}.
\end{equation}

\emph{The Inverse Relative Transfer Function beamformer}, as defined in this work, utilises the RTFs represented by $\g(k,\ell)$ to synchronise the spatial images of the target source on all microphones with the reference one, and sums them together. Therefore,
\begin{equation}
\label{eq:fsb}
\w_{\rm IRTF} = \frac{1}{m}(\g^{-1}),
\end{equation}
where $\g^{-1}$ contains the reciprocal values of the elements of $\g$. 
Compared to MVDR, such processing through IRTF alleviates the need to estimate $\C_{\y\y}$. 
This renders the beamforming operation robust with respect to estimation errors. 
The IRTF could be seen as a generalisation of the Delay-and-Sum beamforming for reverberant environments, because, the DSB and IRTF coincide in free-field conditions. 

The IRTF can also be followed by a single-channel post-filter, such as the Wiener filter described by~(\ref{mmse}). 
The residual noise estimate is obtained as $r_{\rm IRTF}=\w^H_{\rm IRTF}\y$.

\emph{The GEV beamformer} is designed to maximise the signal-to-noise ratio (SNR) in each frequency bin using
\begin{equation}
{\w}_{\rm GEV} = \argmax_{\w} \frac{\wH \C_{\s\s} \w}{\wH \C_{\y\y} \w},
\end{equation}
where $\C_{\s\s}={\rm E}[\s\s^H]$ is the covariance matrix of target speech.
This optimisation problem leads to finding the generalised eigenvector \cite{heymann_chime4}
\begin{equation}
\label{eq:gev}
\C_{\s\s}\w_{\rm GEV} = \lambda \C_{\y\y}\w_{\rm GEV}, 
\end{equation}
where $\lambda$ is the maximal generalised eigenvalue.
Since the norm of $\w_{\rm GEV}$ can be arbitrary, the output of the GEV beamformer has an ambiguous spectrum.
This problem can be solved by applying a single channel post-filter
\begin{equation}
G_{\rm BAN} = \frac{\sqrt{{\w}^H_{\rm GEV}\C_{\y\y}\C^H_{\y\y}{\w}_{\rm GEV}/M}}{{\w}^H_{\rm GEV}\C_{\y\y}{\w}_{\rm GEV}},
\end{equation}
which is called Blind Analytic Normalisation (BAN). 
In theory, the cascade of the GEV beamformer and of the BAN post-filter results in the MVDR beamforming; see, e.g., \cite{umbach_gev}.

\section{Proposed Method} 
\label{sec:method}

\begin{figure*}[t]
    \begin{center}
	\includegraphics[width=0.99\linewidth]{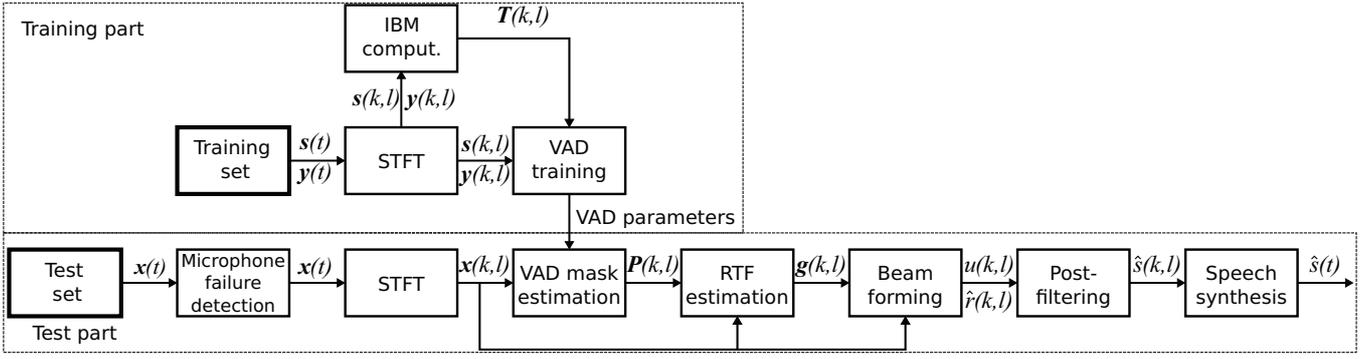}	\caption{\label{fig:blck_diag} Block-diagram of the proposed method. Variables $\s(t),\y(t)\ \text{and}\  \x(t)$ denote time-domain representations of one block of target speech, noise and noisy speech signals, respectively.}
	\end{center}
\end{figure*}

The proposed enhancement algorithm, visualised in Figure~\ref{fig:blck_diag}, consists of the following operations that are applied independently to each block of input signals.
\begin{enumerate}
	\item Detection of microphone failures.
	\item Transformation into time-frequency domain using Short-Time Fourier Transform (STFT).
	\item Estimation of time-frequency masks based on voice activity as detected by pre-trained DNN.
	\item Estimation of RTFs.
	\item Application of selected beamforming technique.
	\item Post-filtering applied to the output of the beamformer.
	\item Synthesis of the speech waveform using the inverse Discrete Fourier Transform and overlap-add.
\end{enumerate}
Two processing regimes are considered: 1) In block-online mode, the signals are 
processed block-by-block where the length of each block is fixed while,
2) in batch mode, the entire input recording is processed as a single block. 

The signal processing part is implemented in Matlab.
The training of the VAD network is performed using the Torch framework~\cite{torch_web} and the Lua language.
Source codes are available on the Internet~\cite{my_asap_web}.

\subsection{Detection of microphone failures} 
\label{sec:failures}

Various hardware malfunctions or measurement errors caused by the user can occur in real-world recordings.
These effects violate the validity of the mixing model \eqref{model}, which can result in a distorted output of the multi-microphone enhancement algorithm.
It is therefore beneficial to detect channels which contain such errors and to skip them for the time interval of the malfunction.

The proposed failure detection proceeds as follows.
For the $i$th microphone signal, $i=1\dots M$, time-domain correlation coefficients with the other channels are computed.
Let the maximum correlation in absolute value be denoted by $\mu_i$.
If $\mu_i$ is smaller than a predefined threshold $t_\mu$, the $i$th channel is discarded from further processing.

\subsection{Time-frequency domain transformation} 
\label{sec:stft}

The STFT is applied to signals sampled at $16$~kHz with a frame length of $512$ samples and a shift of $128$ samples. The frames are weighted by the Hamming window. In the block-online regime, each block consists of $30-250$ STFT frames, which 
corresponds to $0.25 - 2$s.

\subsection{Voice activity detection} 
\label{sec:vad}

The VAD is implemented as a DNN. This type of VAD is able to adapt to complex noisy conditions, provided that enough training data is available.
The network aims to classify which STFT bins are dominated by target speech.

The training data consists of input noisy speech and corresponding target labels, which reflect the true amount of speech in the input signal. The labelling of targets is performed automatically, using a known decomposition of the input signal into the speech component and the background noise.
Our data for VAD training are taken from the simulated part of the CHiME-4~\cite{chime_web} training dataset, i.e., $7138$ utterances are utilised.
A major part of the dataset (80\%) is used to learn the network parameters; the remaining part is used to validate the obtained VADs.

The VAD is proposed for a single channel, that is, the spatial information provided by the multi-channel data is not used.
Henceforth, our detector is denoted as mVAD (multi-output VAD). It performs the detection in each frequency bin separately, so it outputs a vector value whose dimension corresponds with the frequency resolution. 

For practical reasons stemming from the block-online utilisation, the mVAD is designed to have feed-forward fully-connected topology. Although, e.g., the bi-directional long short-term memory (BLSTM) network is able to achieve slightly better detection (see e.g.,~\cite{heymann}), it requires long context of frames to operate. The mVAD is designed to function without any context frames, which allows fully independent processing of subsequent data blocks.

The input layer of mVAD consists of $257$ neurons; the input vector contains spectral magnitudes of the current STFT frame. Next, there are two hidden layers, each containing $1024$ neurons. The activation function between hidden layers is ReLU; the activation preceding the output layer is sigmoid nonlinearity. The output layer consists of $257$ units, which present the VAD decision for each of the input frequency bins.

The mVAD is trained to estimate ideal binary masks (IBM) defined through
\begin{equation}
\label{eq:tm}
T_i^{\text{mVAD}}(k,\ell)=
\begin{cases}
1,&\:10\log\left(\frac{|s_i(k,\ell)|^2}{|y_i(k,\ell)|^2}\right) > t_{\rm SNR}, \\
0,&\:\text{otherwise}.
\end{cases}
\end{equation}
where $i$ is the microphone index, and $t_{\rm SNR}$ denotes a threshold parameter.
During the training, which is finished after $40$ epochs, the Mean Square Error criterion is optimised. 

The mVAD is trained without pre-training, using the gradient descent method.
Prior to the training, the input vectors are normalised to zero mean and unit variance.
The initialisation of both weights and biases are random, drawn from the normal distribution $\mathcal{N}(0,\frac{6}{a+b})$, 
where $a$ and $b$ are the numbers of network's input and output neurons, respectively.
The output of the networks are masks $P_i(k,\ell), i=1\dots M$.
Although the targets $T_i^{\text{mVAD}}$ are binary, the networks output values from $[0,1]$.

As an optional step, the masks $P_i(k,\ell)$ can be condensed into a single mask that is applicable to all channels, which is an operation referred to as pooling. This post-processing step improves the robustness of VAD.
We utilise median pooling, that is, the condensed mask is computed as
\begin{equation}
P(k,\ell)={\rm median}_{i=1\dots M}\, P_i(k,\ell).
\end{equation}

\subsection{Relative Transfer Function Estimation}
\label{sec:rtf}

The proposed RTF estimation is based on a modification of the method from \cite{shalvi1996}. 
This estimator exploits the non-stationarity of speech while assuming that the noise is stationary. 
The latter assumption is somewhat restrictive, since the noise is often non-stationary in real-world situations. 
Our idea for the improvement of this estimator is as follows. 
The estimates of covariance between microphones are weighted by the output of VAD, and thus emphasise bins, which exhibit high signal-to-noise-ratio. In this way, even if the assumption of noise stationarity is not completely true, the contribution of the noisy bins to the RTF estimate is limited and the estimates are computed mainly from bins dominated by speech.

Specifically, let $x_i$, $y_i$ and $g_i^{-1}$ denote the $i$th element of $\x$, $\y$ and $\g^{-1}$, respectively.
Using \eqref{model}, the relationship between the reference and the $i$th channel can be expressed as
\begin{equation}
\label{eq:rtf:model}
x_{\rm ref}(k,\ell) = g_i^{-1}(k) \cdot x_i(k,\ell) + v_i(k,\ell)
\end{equation}
where $v_i(k,\ell) = y_{\rm ref}(k,\ell) - g_i^{-1}(k) \cdot y_i(k,\ell)$. 
Now, let each processing block be divided into $N$ equally long sub-blocks. 
Based on \eqref{eq:rtf:model}, the (cross-)Power Spectral Densities of the channels within the $n$th sub-block satisfy
\begin{equation}
\phi_{x_{\rm ref},x_i}(k,n) = g_i^{-1}(k) \phi_{x_i,x_i}(k,n) + \phi_{v_i,x_i}(k,n),
\end{equation}
where $\phi_{v_i,x_i}(k,n)$ is independent of $n$ when the noise is stationary within the block \cite{shalvi1996}. 
By substituting the PSDs with their sample-based estimates (denoted by $\widehat{\phi}$), $N$ linear equations are obtained (one for each sub-block) 
\begin{equation}
\label{eq:statio:est}
\widehat{\phi}_{x_{\rm ref},x_i}(k,n) 
= g_i^{-1}(k)\widehat{\phi}_{x_i,x_i}(k,n) + \phi_{v_i,x_i}(k,n)
+ \epsilon(k,n),
\end{equation}
where $\epsilon(k,n)$ is the estimation error
\begin{equation}
\epsilon(k,n) = \widehat{\phi}_{v_i,x_i}(k,n) - \phi_{v_i,x_i}(k,n).
\end{equation}

Let the sub-block length be denoted by $L_0$. The weighted \mbox{(cross-)channel} PSD estimates within the $n$th sub-block are 
\begin{align}
	\label{eq:PSDestim}
	\widehat{\phi}_{x_{\rm ref},x_i}(k,n) &= \sum_{\ell=nL_0}^{(n+1)L_0-1} P_i(k,\ell) x_{\rm ref}(k,\ell) x_i^{*}(k,\ell) \nonumber \\
	\widehat{\phi}_{x_i,x_i}(k,n) &= \sum_{\ell=nL_0}^{(n+1)L_0-1} P_i(k,\ell) x_i(k,\ell) x_i^{*}(k,\ell),
\end{align}
where $\cdot^*$ denotes conjugation.
By setting $P_i(k,\ell)=1$, the weighting of the frames by VAD output is disabled and the RTF estimation coincides with the original method from~\cite{shalvi1996}.

Now, formula \eqref{eq:statio:est}, where $n=0,\dots,N-1$, gives an over-determined system of linear equations with two unknown variables $g_i^{-1}(k)$ and $\phi_{v_i,x_i}(k,n)$. 
As suggested in \cite{shalvi1996}, the least-squares solution gives us the final estimates of these quantities. 
They can be computed using the closed-form expression (the arguments $k$ and $n$ are omitted)
\begin{equation}
\label{eq:rtf:est}
\widehat{g}_i^{-1} = \frac{
	\left\langle \widehat{\phi}_{x_{\rm ref},x_i}\widehat{\phi}_{x_i,x_i} \right\rangle 
	- \left\langle \widehat{\phi}_{x_{\rm ref},x_i}\right\rangle \left\langle \widehat{\phi}_{x_i,x_i} \right\rangle}
{ \left\langle \widehat{\phi}^2_{x_i,x_i} \right\rangle - \left\langle \widehat{\phi}_{x_i,x_i} \right\rangle^2},
\end{equation}
where $\left\langle \cdot \right\rangle$ denotes the averaging operator over the sub-block index $n$. 

\subsection{Beamforming: Implementation details}

To implement the MVDR beamformer \eqref{mvdr} in practice, an estimate of the noise covariance $\C_{\y\y}$ must be available. 
Since the RTF estimate has already been computed, we proceed by constructing a blocking matrix that is orthogonal to $\widehat{\g}$; 
it blocks the target signal and outputs a noise reference
\begin{equation}
\label{eq:mvdr:noise}
\vv = \B\x.
\end{equation}
 
The blocking matrix $\B$ can have the structure
\begin{equation}
\label{eq:bmatrix}
{\B}=
\begin{bmatrix}
-1 & g^{-1}_2 & 0 & \dots & 0\\
-1 & 0 & g^{-1}_3 & \dots & 0\\
\vdots & \vdots & \vdots & \ddots & \vdots \\
-1 & 0 & 0 & \dots & g^{-1}_M\\
\end{bmatrix},
\end{equation}
where it is assumed, without any loss of generality, that the reference channel $x_{\rm ref}$ corresponds to the first microphone.

To obtain the estimate of noise signals $\y$ as they appear in mixture~(\ref{model}), 
the least square estimate \cite{koldovsky2016} is applied
\begin{equation}
\widehat{\y} = {\C_{\x\x}}{\B}^H({\B}{\C_{\x\x}}{\B}^H)^{-1}{\B}{\vv}, 
\label{noiseest}
\end{equation}
where $\C_{\x\x}={\rm E}[\x\x^H]$ is replaced by its sample-based estimate.
The covariance of $\widehat{\y}$ is equal to
\begin{equation}
{\C}_{\hy\hy}={\rm E}[\widehat{\y}\widehat{\y}^H]={\C_{\x\x}}{\B}^H({\B}{\C_{\x\x}}{\B}^H)^{-1}{\B}{\C_{\x\x}},
\end{equation}
which is substituted into (\ref{mvdr}). 
However, ${\C}_{\hy\hy}$ is rank deficient having the rank $\leq M-1$; therefore, its inverse matrix does not exist. As proposed in \cite{koldovsky2016}, the matrix ${\C}_{\y\y}^{-1}$ is replaced by the Moore-Penrose pseudoinverse of ${\C}_{\hy\hy}$, denoted as ${\C}_{\widehat{\y}\widehat{\y}}^{\dagger}$. The implementation of MVDR is thus finally given by
\begin{equation}\label{MVDR1}
\widehat{\w}_{\rm MVDR}=\frac{{\C}_{\widehat{\y}\widehat{\y}}^{\dagger}\widehat{\g}}{\widehat{\g}^H{\C}_{\widehat{\y}\widehat{\y}}^{\dagger}\widehat{\g}}.
\end{equation}

In the case of the IRTF beamformer, the implementation is straightforward through formula~(\ref{eq:fsb}) where the true RTF is replaced by its estimated value. 

\subsection{Single-channel post-filtering}
\label{sec:post}

The single-channel post-filtering is applied with the aim to minimise the 
residual noise in the beamforming output. 
In our work, the residual noise is estimated as 
\begin{align}
	\widehat{r}_{\rm MVDR}=&\widehat{\w}^H_{\rm MVDR} \widehat{\y},\nonumber \\
	\widehat{r}_{\rm IRTF}=&\widehat{\w}^H_{\rm IRTF} \widehat{\y},
\end{align}
where $\widehat{\y}$ is obtained using \eqref{eq:bmatrix} and \eqref{noiseest}.
Assuming that the beamforming output $u(k,l)$ consists of the target speech 
signal and of the residual noise, the approximate Wiener mask is applied defined as 
\begin{equation}
\label{eq:wiener:impl}
G(k,\ell)=\frac{\max\{|u(k,\ell)|^2-|\widehat{r}(k,\ell)|^2,\delta\}}{|u(k,\ell)|^2+\delta},
\end{equation}
where $\delta$ is a small positive constant to prevent from division by zero.
Finally, the estimate of the target speech is given by 
\begin{equation}
\label{eq:wiener:post}
\widehat{s}(k,\ell) = G(k,\ell)u(k,\ell).
\end{equation}

We add the following three heuristic modifications of \eqref{eq:wiener:impl} 
based on practical experience to improve the performance of this post-filtering 
step.
\begin{enumerate}
	\item For frequencies lower than $f_{\rm min}$~Hz, $G(k,\ell)=0.01$. 
	As observed on CHiME-4 data, the very low frequencies usually contain only noise, so it is better to attenuate them.
	\item For higher frequencies than $f_{\rm max}$~Hz, $G(k,\ell)=1$. This is 
	due to the fact that a speech signal is more difficult to block in the 
	higher frequency band, as there is typically a lower SNR. The RTF estimate 
	has a higher variance, which causes leakage of the target signal into the 
	residual noise estimate, and finally a distortion due to the post-filter. 
	
	\item We set $G(k,\ell)=1$ when $P(k,\ell)>t_{\rm VAD}$. The goal is to 
	preserve time-frequency regions where VAD detects speech with sufficiently 
	high probability. 
\end{enumerate}

To summarise, the proposed  method is described in Algorithm~\ref{alg:summary}. Specific choices of the parameters that are used in experiments are listed in Table~\ref{tab:params}.

\begin{algorithm}
	\caption{Enhancement of a processing block by the proposed method}
	\label{alg:summary}
	\begin{algorithmic}
		\REQUIRE Multi-channel speech signal
		\STATE Discard channels with microphone failures;
		\STATE Apply STFT to input speech to compute $\x(k,\ell)$;
		\FOR[excluding reference channel 1]{$i\leftarrow 2$ \TO $M$}		
			\STATE Compute VAD mask $P_i(k,\ell)$ using $x_i(k,\ell)$, mVAD;
			\COMMENT {Section~\ref{sec:vad}}	
			\STATE Estimate RTF $g_i(k)$ using $P_i(k,\ell)$ and $x_i(k,\ell)$;			
			\COMMENT {Equations~\eqref{eq:PSDestim} and~\eqref{eq:rtf:est}}	
		\ENDFOR
		\STATE Compute beamforming output $u(k,\ell)$ using either \\ IRTF~\eqref{eq:fsb} or MVDR~\eqref{eq:mvdr:noise}-\eqref{MVDR1};		
		\STATE Apply post-filter~\eqref{eq:wiener:impl}-\eqref{eq:wiener:post} to gain target $\hat{s}(k,\ell)$;				
		\STATE Reconstruct the enhanced speech in time-domain;	
	\end{algorithmic}
\end{algorithm}

\section{Experimental results and discussion}
\label{sec:experiments}

\begin{table}
    \begin{minipage}{0.47\textwidth}
	\caption{\label{tab:params} The values of the free parameters used in the experiments. The values were selected based on preliminary experiments with the CHiME-4 data. The sub-block number $N$ is selected, such that the sub-block always contains $10$ frames ($80$~ms of speech).}
	\begin{center}
	\begin{tabular}{|c|c|c|}
		\hline
		\textbf{Param.} & \textbf{Value} & \textbf{Meaning} \\
		\hline
		$M$ & $5$ or $6$ & Number of channels: CHiME-4\\
		& $4$ & Number of channels: Dynamic dataset\\		
		\hline
		$N$ & 		  				& Sub-block number, RTF estim. 
		(see~\ref{sec:rtf}) \\
		& $3,5,10,25 $ & For block lengths $0.25, 0.4, 0.8, 2$ s  \\
		\hline
		$t_{\rm SNR}$ & $5$dB & Threshold, mVAD (see~\eqref{eq:tm}) \\
		\hline
		$f_{\rm max}$ & $3125$~Hz & Threshold, Wiener filter 
		(see~\ref{sec:post})\\
		\hline
		$f_{\rm min}$ & $100$~Hz & Threshold, Wiener filter 
		(see~\ref{sec:post})\\
		\hline
		$t_{\rm VAD}$ & $0.3$ & Threshold, Wiener filter (see~\ref{sec:post})\\
		\hline
		$t_\mu$ & 		  & Threshold, mic. failure detect. 
		(see~\ref{sec:failures})\\
		& $0.05$ & CHiME-4 simulated, Dynamic \\
		& $0.40$ & CHiME-4 real-world datasets\\
		\hline
	\end{tabular}
	\end{center}
	\end{minipage}
\end{table}

This section presents experiments conducted on two datasets; 
the CHiME-4 dataset~\cite{chime_web} and another dataset featuring moving speaker, which is further denoted as "Dynamic".
On both datasets, we perform objective evaluation using the criteria defined in the 
BSS\_Eval toolbox~\cite{bsseval} and the evaluation of perceptual quality of 
enhanced signals using PESQ~\cite{pesq}; the wide-band ($16$~kHz) implementation from~\cite{loizou} is used. The PESQ measure yields high correlation coefficient with overall quality and enhanced signal distortion obtained using subjective listening tests~\cite{pesq_subj_corr_loizou}.
For CHiME-4, the availability of a trained ASR system and of labelled test utterances is taken into account. The multi-channel enhancement is used to improve the performance of automatic speech recognition in terms of WER.
In the experiments, the influence of various processing steps and of short duration of the input signal on the performance is analysed.

CHiME-4 defines three tracks where utterances of speaking persons are recorded 
in noisy conditions; our experiments focus on the multi-channel track with six-channel 
recordings. These were either simulated (SIMU) or acquired by a tablet device 
in real-world situations (REAL). With the REAL recordings, we do not use microphone 2, because it is oriented in a direction away from the speaker. For SIMU test set we use another experimentally determined subset of channels, namely 2 to 6. 
The dataset contains data from four different noisy environments (bus, cafeteria, 
street, and pedestrian area). The utterances are 
taken from the WSJ corpus \cite{wsj0} and are provided along with the 
respective transcription. 
There are $5920$ test files overall, which are divided into development and test datasets.

In the Dynamic dataset, the signal part corresponds to a $59$ seconds long utterance recorded on $4$ microphones in a highly reverberated room ($T_{60}\approx 700$~ms). The speaker was sitting about $0.5$~m in front of the microphone array and was leaning its upper body to sides by about $25$~cm to the left or right from the central position. 
As the noise part of the data, the noise types from the CHiME-4 campaign were used (bus, cafeteria, street and pedestrian area), because the utilised VADs were trained for these environments. For the detailed analysis, the speech was mixed with noise at global SNR of $5$~dB. We present also additional experiments, where the influence of the input SNR is studied; here the SNR varies from $0-8$~dB. Channels $1,3,4,6$ of the six-channel track are used. There are thus four noisy instances of the recorded speech, so the total length of the Dynamic dataset is about $4$~minutes.

To perform the objective evaluation, the ground-truth clean speech 
signals are needed. Therefore, we present the objective evaluation only for the 
SIMU development dataset of the CHiME-4 and of the Dynamic dataset, where the reference data is available.
BSS\_Eval is applied by using the \texttt{bss\_decomp\_tvfilt} function for 
signal decompositions. This function utilises a time-varying filter (we set its 
length to $32$ taps) as an allowed distortion of the estimated target source.
The signal-to-interference (SIR), signal-to-distortion (SDR) and 
signal-to-artefact (SAR) ratios are subsequently computed as defined in~\cite{bsseval}.

Speech recognition is performed for the CHiME-4 datasets, where the text reference is available, by the baseline transcription system provided by CHiME-4 
organisers~\cite{chime_data_soft}.
The recogniser features a hybrid DNN-Hidden Markov Model for acoustic modelling and a 5-gram language model. Subsequently, rescoring of the lattice is performed using a recurrent neural 
network language model (RNNLM).

The performance of the proposed system has been compared with two state-of-the-art techniques that were successfully used to solve the CHiME-4 challenge, namely 1) the baseline method BeamformIt \cite{beamformit} 
and 2) the GEV beamformer that is the front-end part of the system described in \cite{heymann_chime4, heymann}.
{These methods represent two frequently used approaches for beamformer steering in reverberant environment: the compensation of time-difference of arrival and the eigenvector decomposition of speech covariance matrix, respectively.} 
While BeamformIt operates only in the batch regime (thus we apply it only to more static CHiME-4 data) as it passes twice through each recording in order to optimise its inner parameters, GEV is also modified for the block-online processing.

The training procedure for the VAD in GEV was kindly 
provided to us by its authors \cite{heymann}. This VAD was retrained using the 
CHiME-4 training datasets. 
The fully-connected variant of the VAD (denoted as hVAD) is utilised, because it possesses comparable topology to that of mVAD.

For the sake of clarification, the compared systems will be denoted by $A$:$B$ 
where $A$ denotes the beamforming method, and $B$ denotes the particular 
modification focused on by the experiment. For example, when VAD selection is 
discussed, the labels could be IRTF:mVAD, IRTF:hVAD, etc.
The presented results are averaged over the four noisy environments; the 
resulting WER reductions are absolute. Also, see Table~\ref{tab:params} for 
other values of parameters utilised in the experiments.

\subsection{Block-online vs. batch processing}
\label{chap:batch}

This experiment is focused on the performances of the compared methods as they 
depend of the length of processing block. The proposed systems utilise mVAD, 
pooling of VAD masks, and the Wiener post-filtering. 
The GEV beamformer is applied with the BAN post-filtering.

\subsubsection{CHiME-4 datasets}

\begin{figure}
    \begin{center}
	\includegraphics[width=0.49\linewidth]{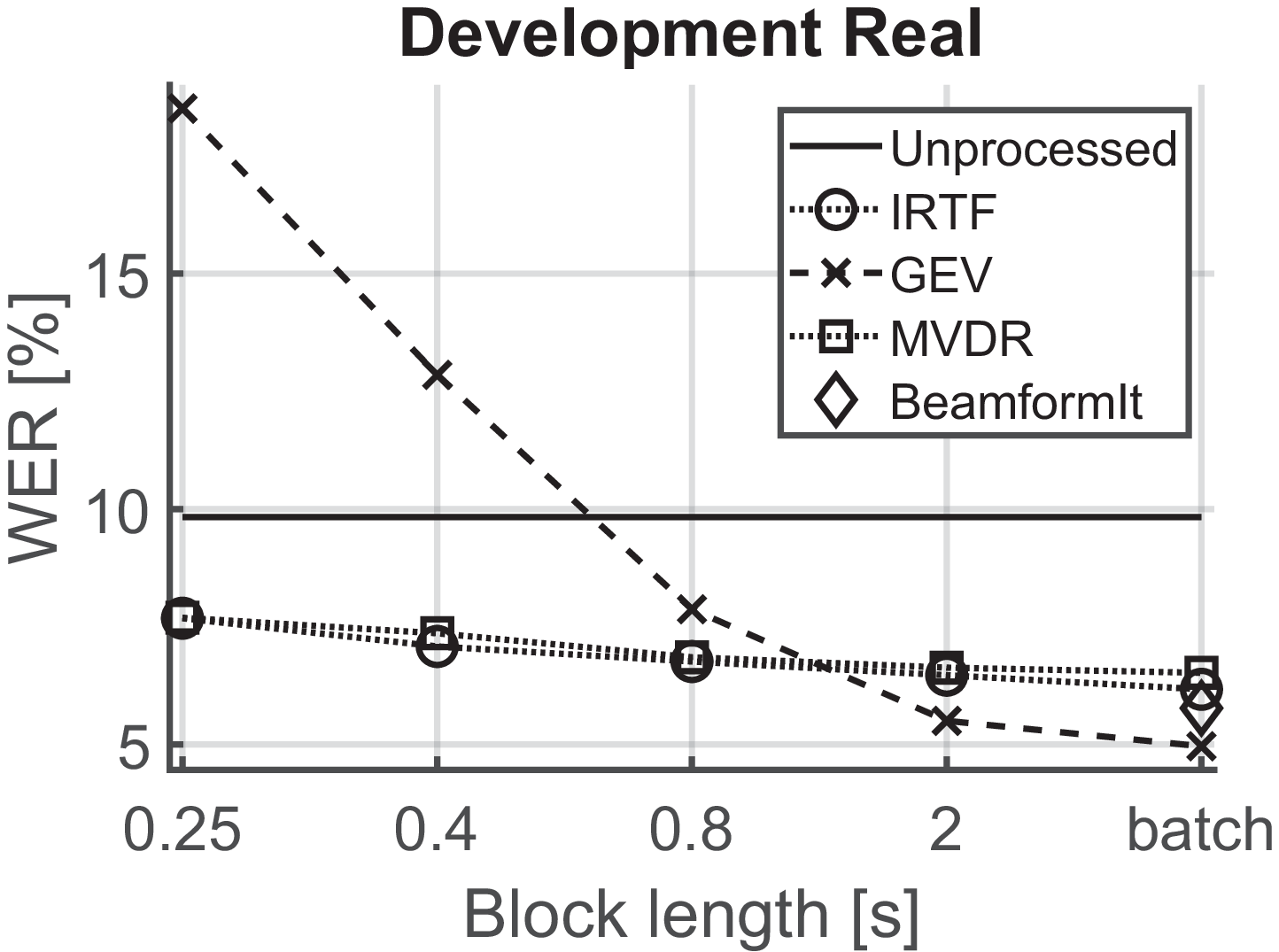}
	\includegraphics[width=0.49\linewidth]{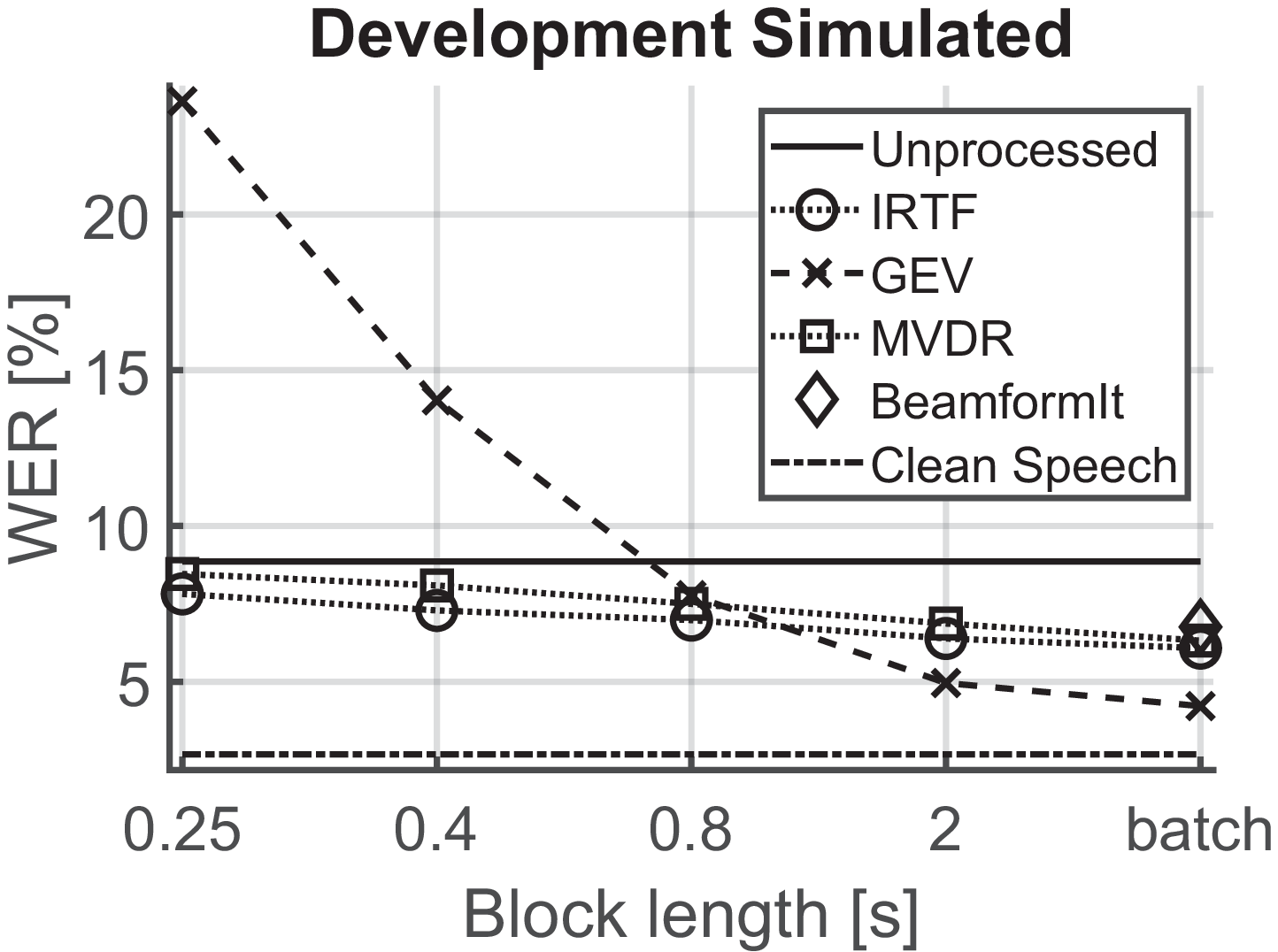}
	\includegraphics[width=0.49\linewidth]{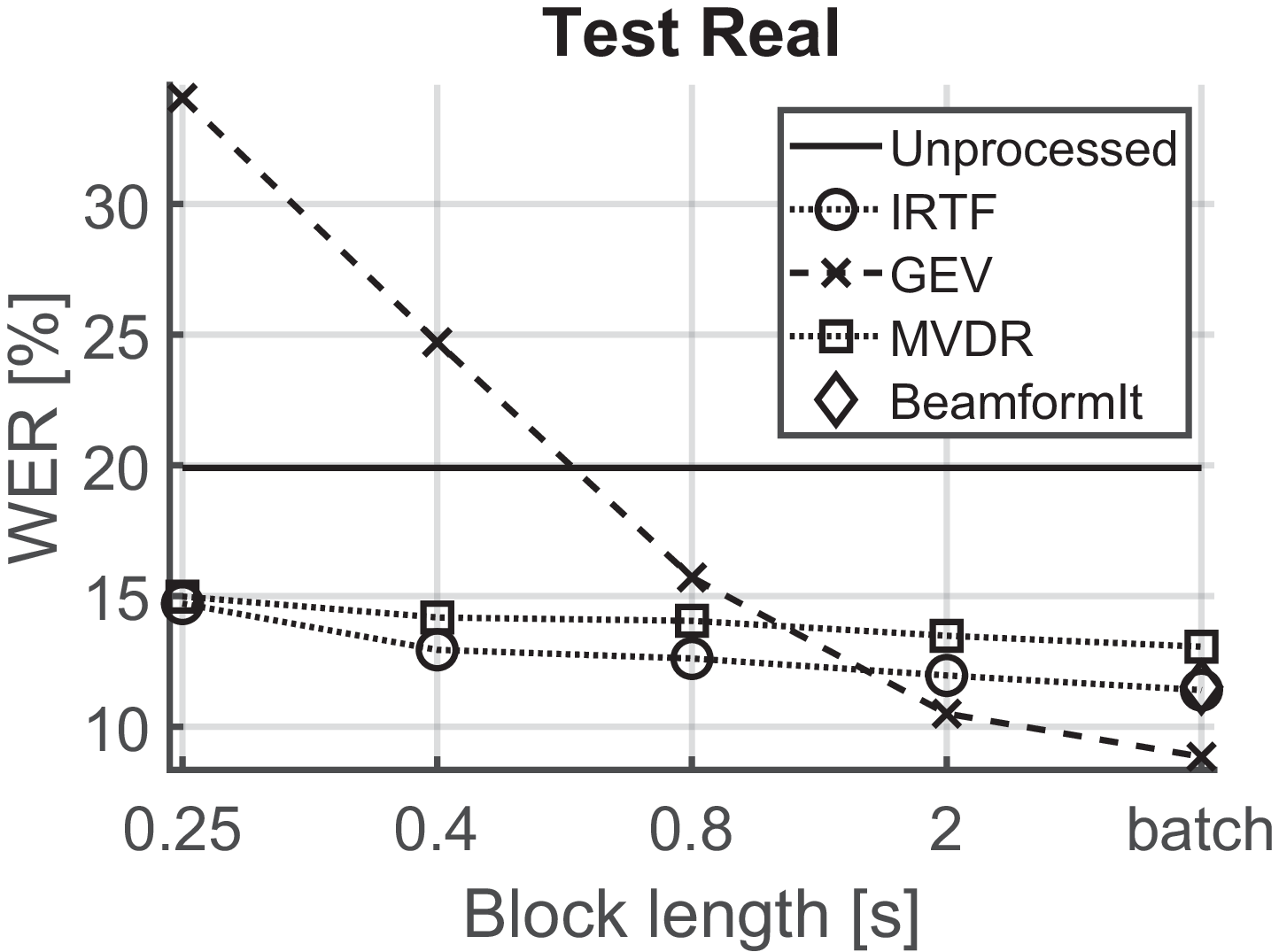}
	\includegraphics[width=0.49\linewidth]{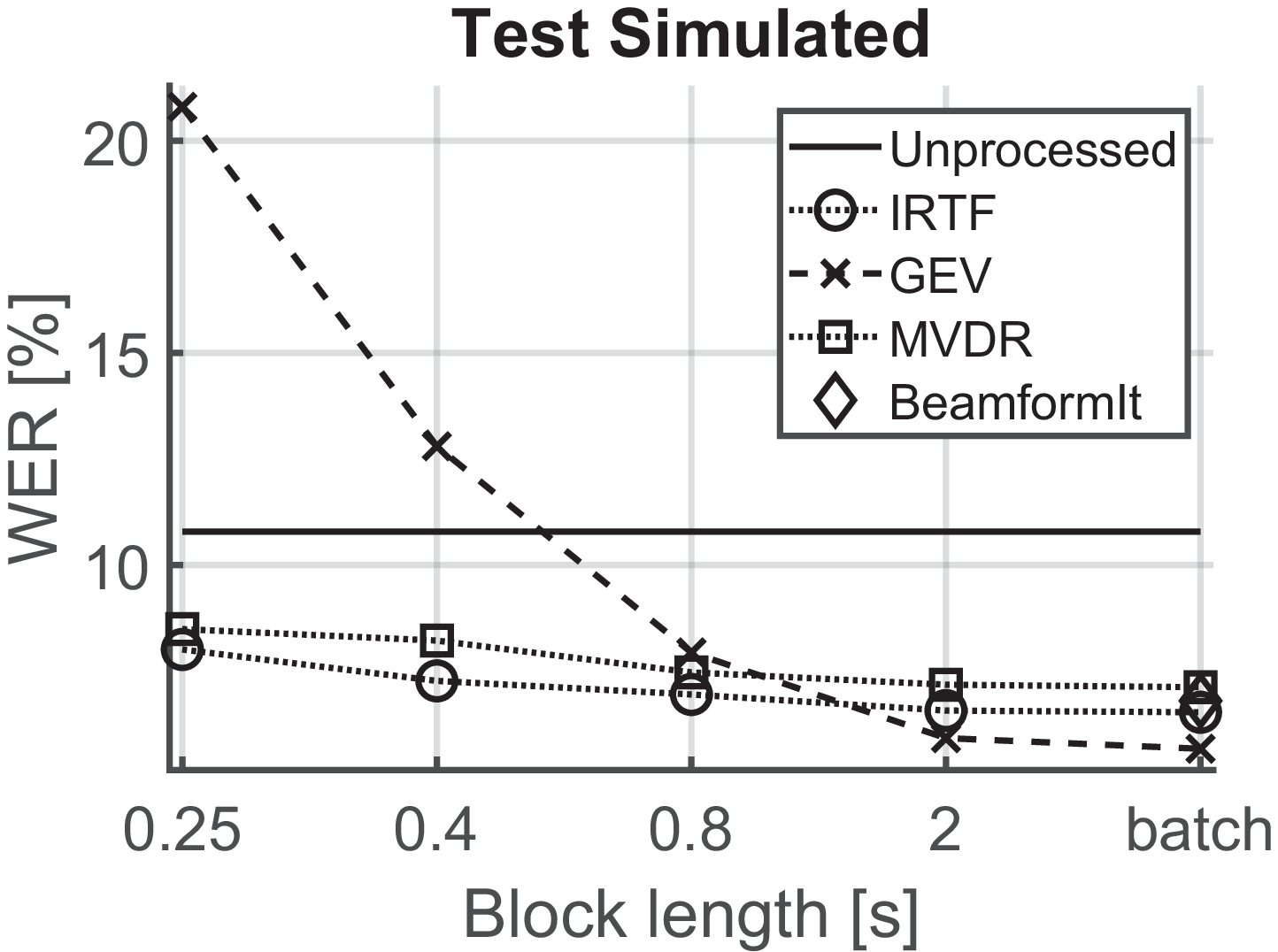}
	\caption{\label{fig:wer_len} CHiME-4: WER [\%] as a function of processing block 
		length. The proposed beamformers use mVAD detector and mask pooling.
		WER of clean speech is available only for the development simulated set 
		due to the rules of CHiME-4.}
    \end{center}
\end{figure}

\begin{figure}
    \begin{center}
	\includegraphics[width=0.49\linewidth]{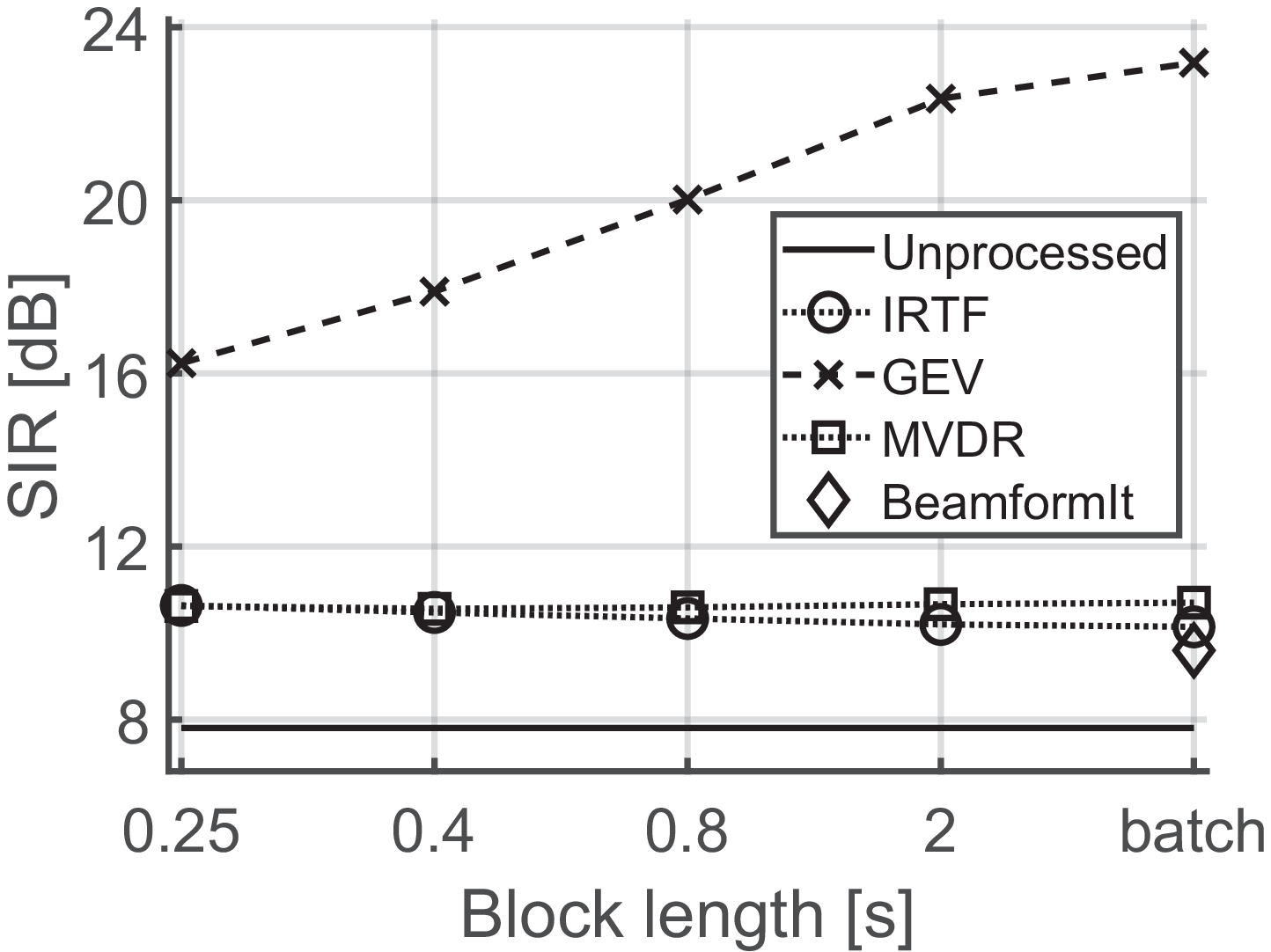}
	\includegraphics[width=0.49\linewidth]{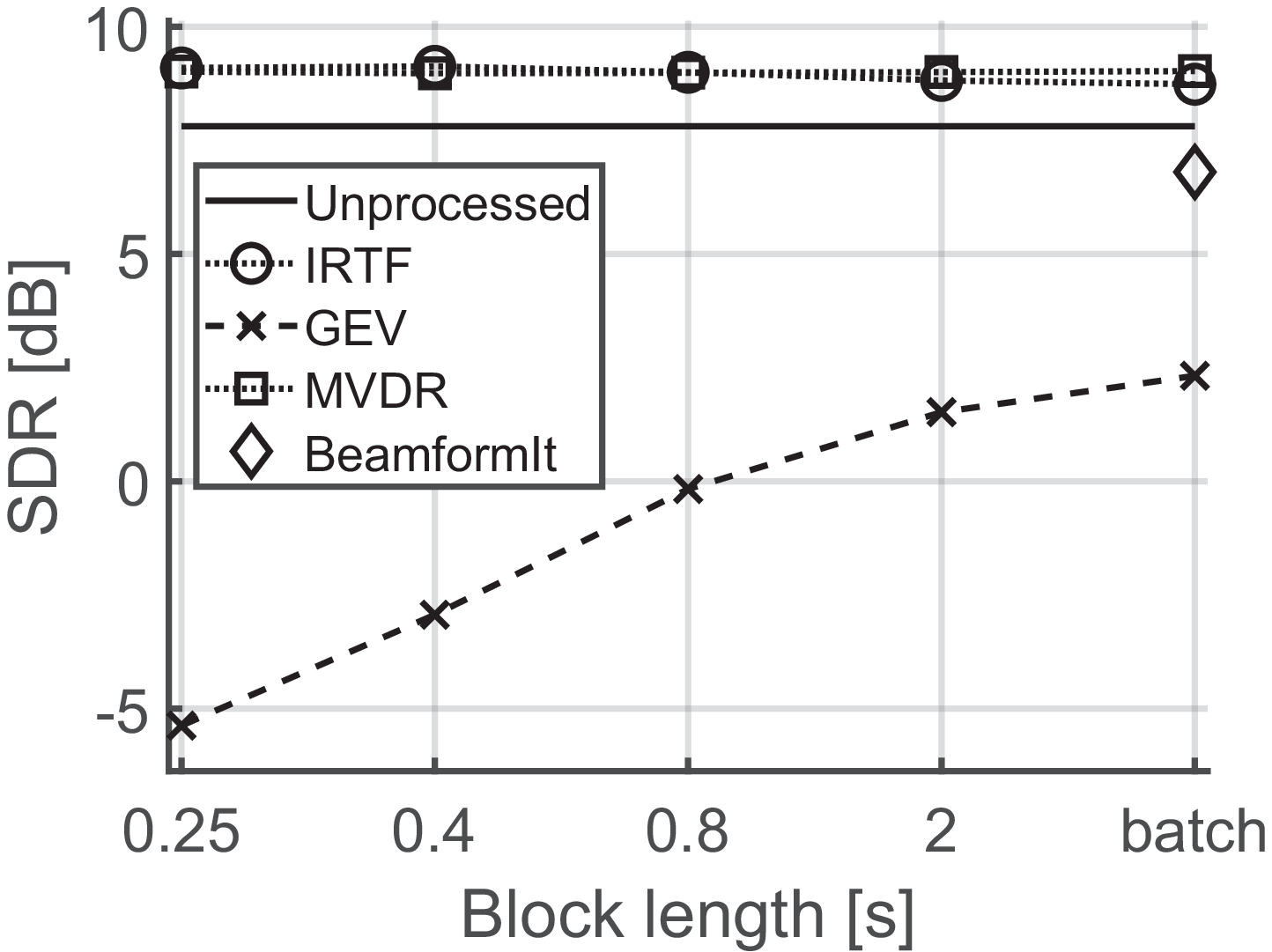}
	\includegraphics[width=0.49\linewidth]{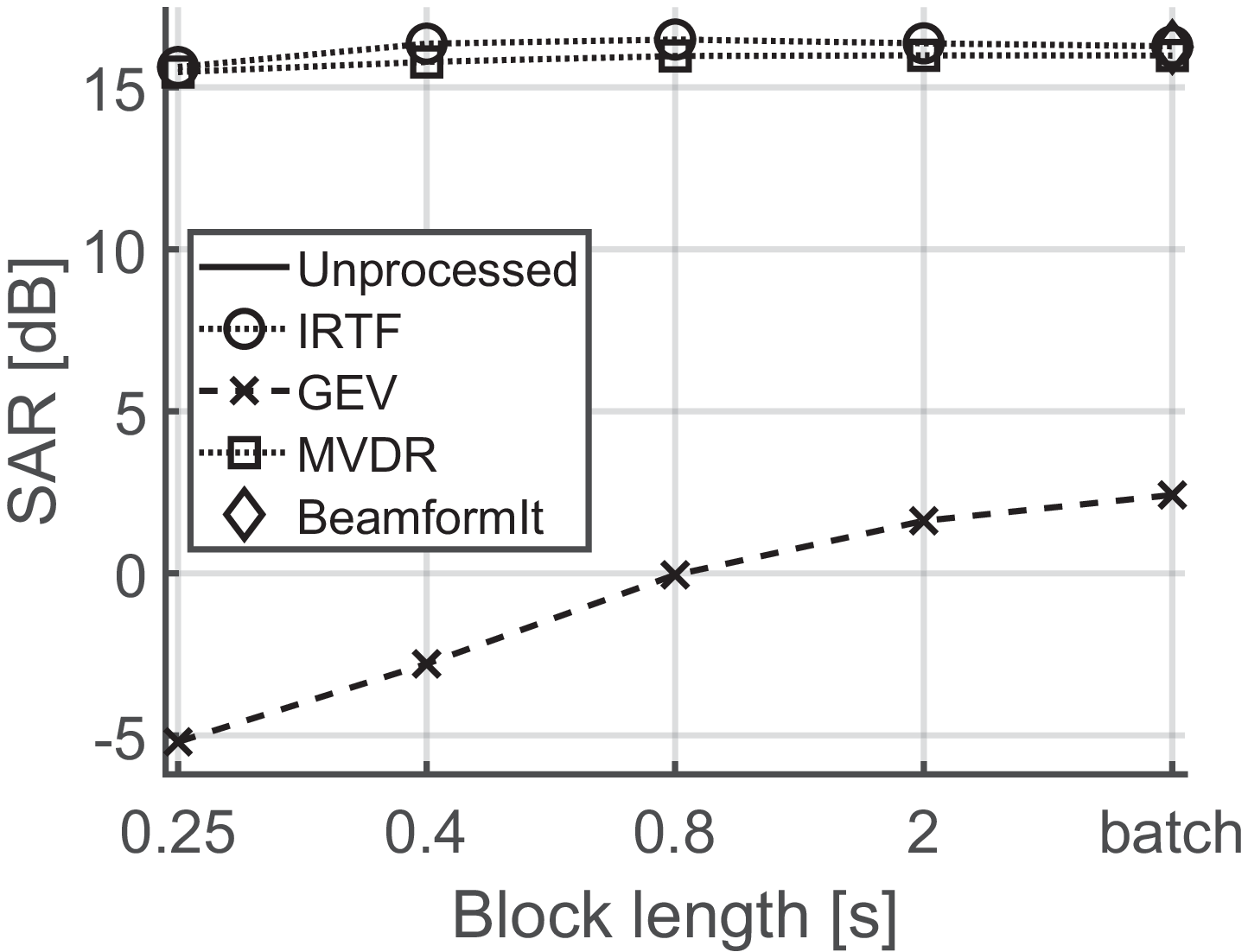}
	\includegraphics[width=0.49\linewidth]{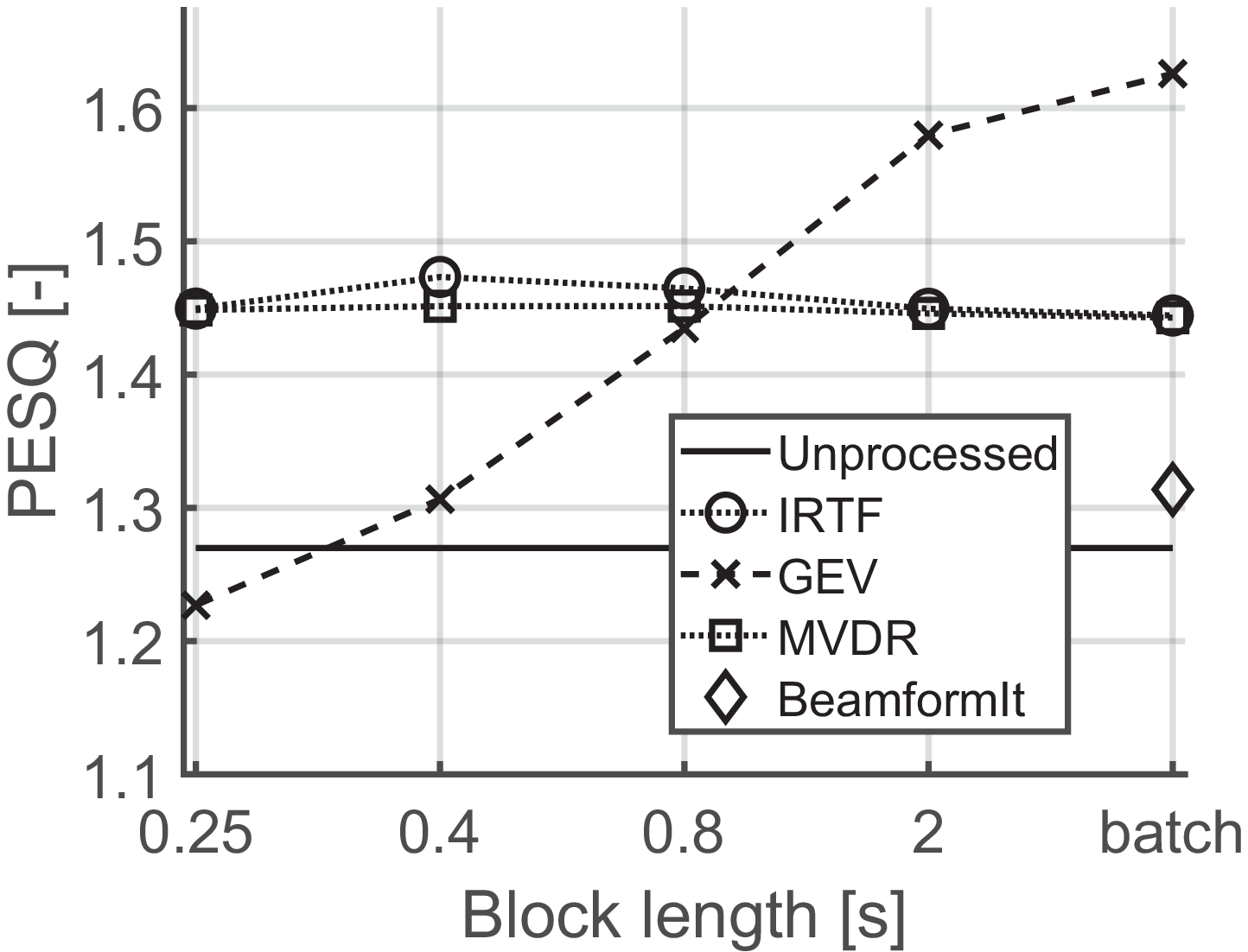}
	\caption{ \label{fig:bss_len} CHiME-4, development SIMU: Objective criteria and PESQ as functions of processing block length. The proposed beamformers use mVAD detector and 
		mask pooling. BeamformIt is presented only for the batch processing mode. SAR 
		of the unprocessed signals is infinite, therefore, it is not shown.}
	\end{center}	
\end{figure}

The results are shown in Figures \ref{fig:wer_len} and \ref{fig:bss_len}. They 
clearly show that all the methods improve their ability to suppress the 
background noise with growing length of the processing blocks, and achieve 
their optimum performance in the batch processing mode. The latter observation  
confirms the fact that the target speaker has, in CHiME-4 recordings, an almost 
fixed position, because otherwise the batch processing would have failed.

The proposed methods yield stable performance with respect to the length of the 
processing block. IRTF achieves comparable or slightly lower WER (by up to 
$1.5$\%) compared to MVDR. GEV:batch achieves the best WER and surpasses 
IRTF:batch by $0.8-2.2$\% WER. BeamformIt yields slightly better results 
compared to IRTF:batch on real datasets and slightly worse results by $0.3-0.7$\% on 
the simulated ones.

The performance of GEV drops down with the decreasing length of the processing block, 
especially, when this goes below $2$ seconds. 
This sensitivity seems to be mainly caused by the increased amount of artefacts 
in the enhanced speech, as shown by SDR and SAR in 
Fig.~\ref{fig:bss_len}. We conjecture that this is partly due to less precise estimates of speech/noise covariance matrices and partly due to the limitations of  
BAN as stated in \cite{umbach_gev}. {BAN relies on a precise 
	estimation of $\w_{\rm GEV}$ and assumes that the norm of $\g(k,\ell)$ is 
	approximately constant across frequencies. Using short blocks of signals, these 
	quantities are estimated with higher estimation errors, and BAN seems to be 
	ineffective.}
Compared to unprocessed speech, PESQ (ranging from $5$ 
	 for the best quality to $1$  for the worst) achieved by GEV:batch is increased by about $0.4$, whereas GEV:$0.25$s lowers the PESQ by about $0.05$. This performance decrease for short processing blocks could be probably alleviated through recursive implementation of GEV. This would, however, decrease the adaptation speed of the technique. 

Compared to the proposed solutions, GEV achieves strong 
noise suppression measured by SIR. For example, GEV:$0.8$s outperforms 
IRTF:$0.8$s by about $9$dB in terms of SIR. However, SDR and SAR obtained by 
GEV:$0.8$s are lower by about $9$ and $16$dB, respectively, compared to 
IRTF:$0.8$s.

Concerning the functionality with respect to noise type: all the compared methods are successful in suppressing the cafeteria, street and pedestrian area noises to a certain extent. For the batch processing, the difference in WER for these environments is up to $3\%$ for the FSB beamformer and up to $2\%$ for GEV. A lower performance is achieved for the bus environment (especially in the REAL task); WER is higher by about $6-9\%$ compared to the other environments for GEV and FSB, respectively. 
We conjecture that the deterioration arises because the bus noise has a narrow-band character concentrated in low frequencies (up to $1$~kHz). These frequencies (above $250$~Hz) are important to ASR because they contain important vowel formants. However, the low-frequency noise is difficult to suppress for the beamformers, because of the long wavelengths of the sound.

\subsubsection{Dynamic dataset}

\begin{figure}
    \begin{center}
	\includegraphics[width=0.49\linewidth]{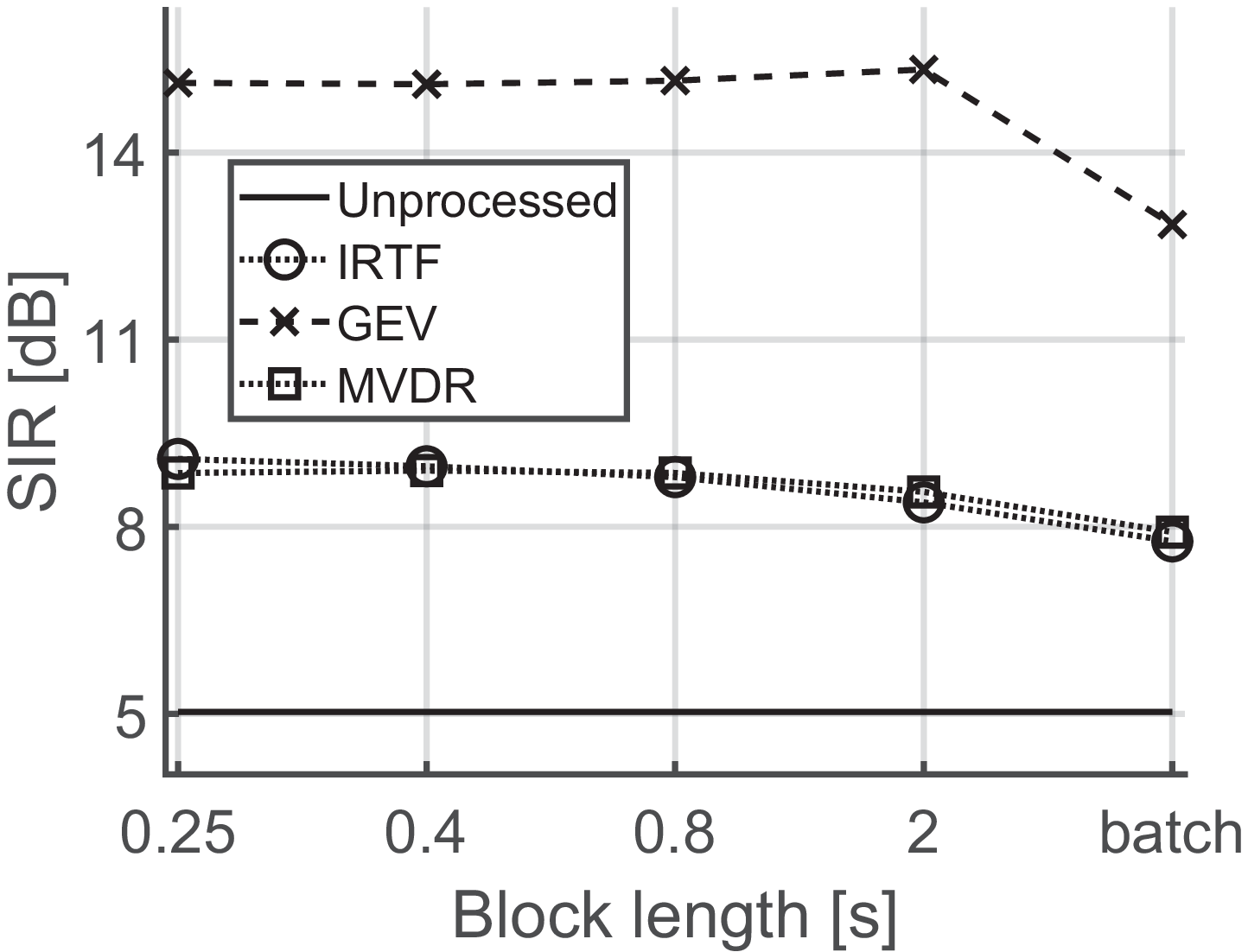}
	\includegraphics[width=0.49\linewidth]{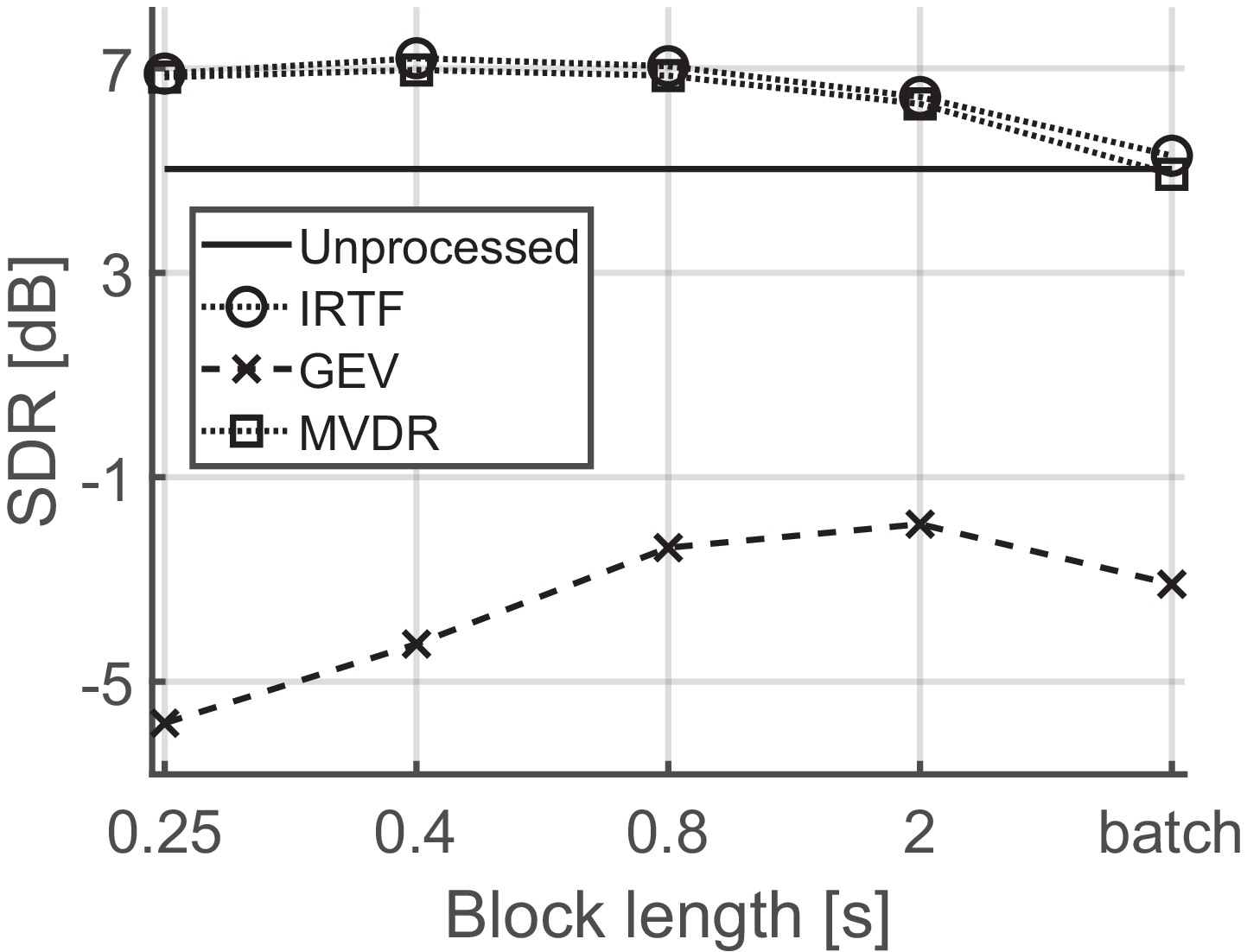}
	\includegraphics[width=0.49\linewidth]{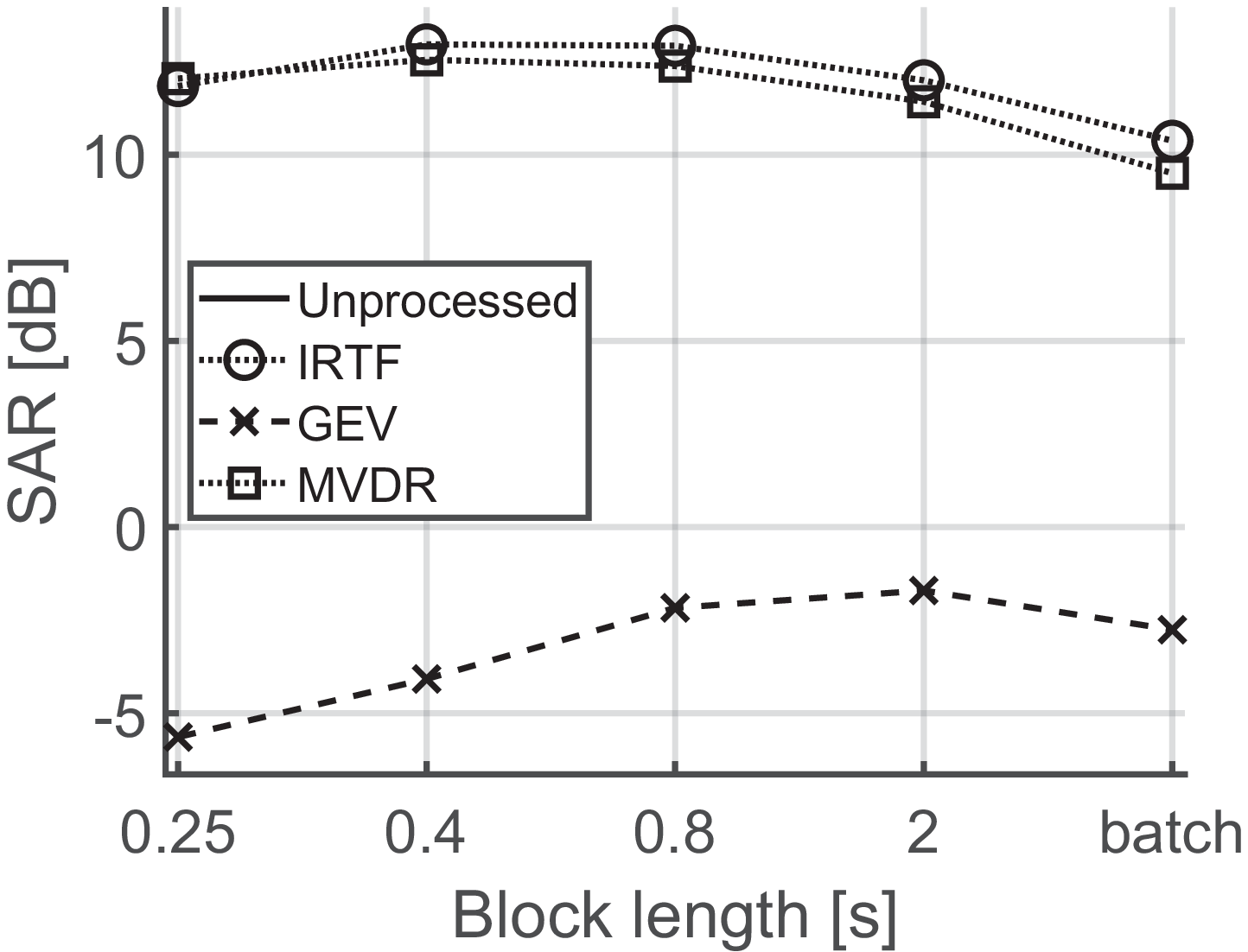}
	\includegraphics[width=0.49\linewidth]{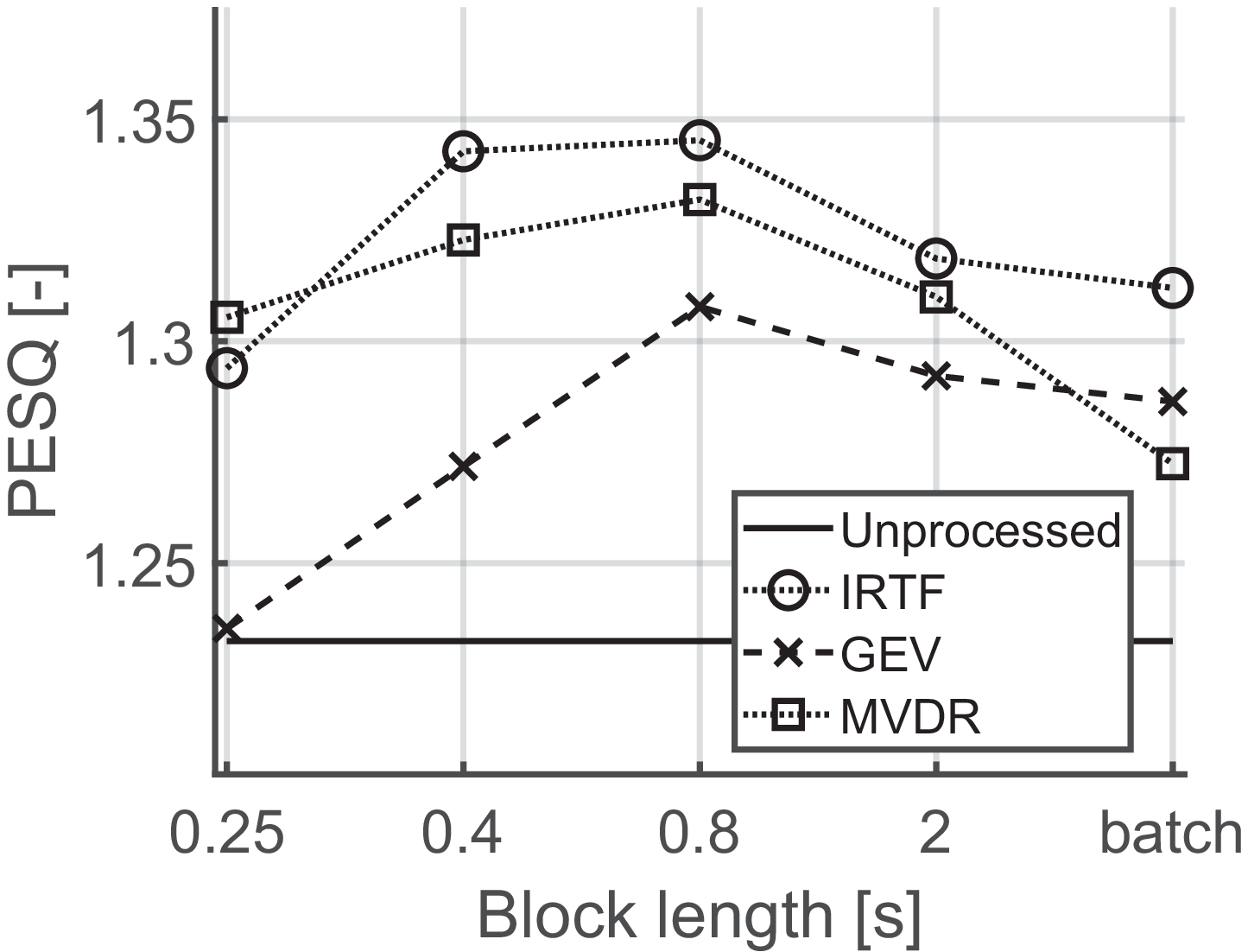}
	\caption{ \label{fig:bss_len_dyn} Dynamic: Objective criteria and PESQ as functions of 
		processing block length. The proposed beamformers use mVAD detector and 
		mask pooling. SAR of the unprocessed signals is infinite, therefore, it is not shown.}
	\end{center}	
\end{figure}

The results in Fig.~\ref{fig:bss_len_dyn} suggest that the most suitable block length for the Dynamic dataset is $0.8$~s. With increasing block duration up to $0.8$~s, the compared methods exhibit improvements in most of the observed metrics. In contrast to CHiME-4 datasets, most of the metrics deteriorate when the block size exceeds $2$~s. This indicates that the signal and noise spatial statistics change in time due to the movements of the speaker.

For processing block lengths $0.8$~s and lower, the obtained results are consistent with our findings reported on CHiME-4 datasets. The proposed methods yield stable performance in terms of SIR, SDR and SAR. In contrast, the performance of the GEV beamformer deteriorates due to the presence of distortions and artefacts in the processed speech, as is indicated by the SDR and SAR metrics. However, GEV retains its strong noise suppression ability indicated by high SIR values.

\begin{figure}
    \begin{center}
	\includegraphics[width=0.49\linewidth]{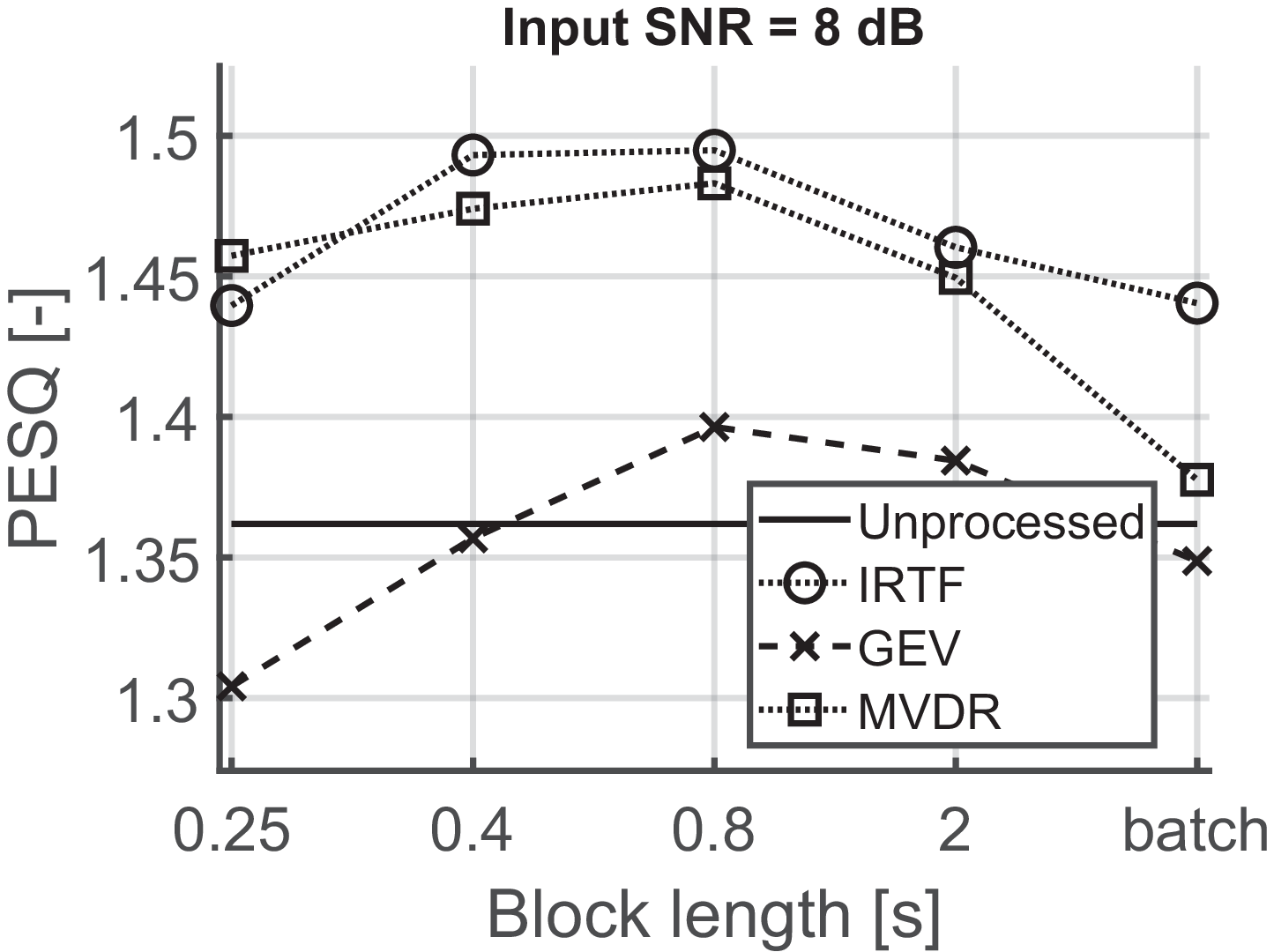}
	\includegraphics[width=0.49\linewidth]{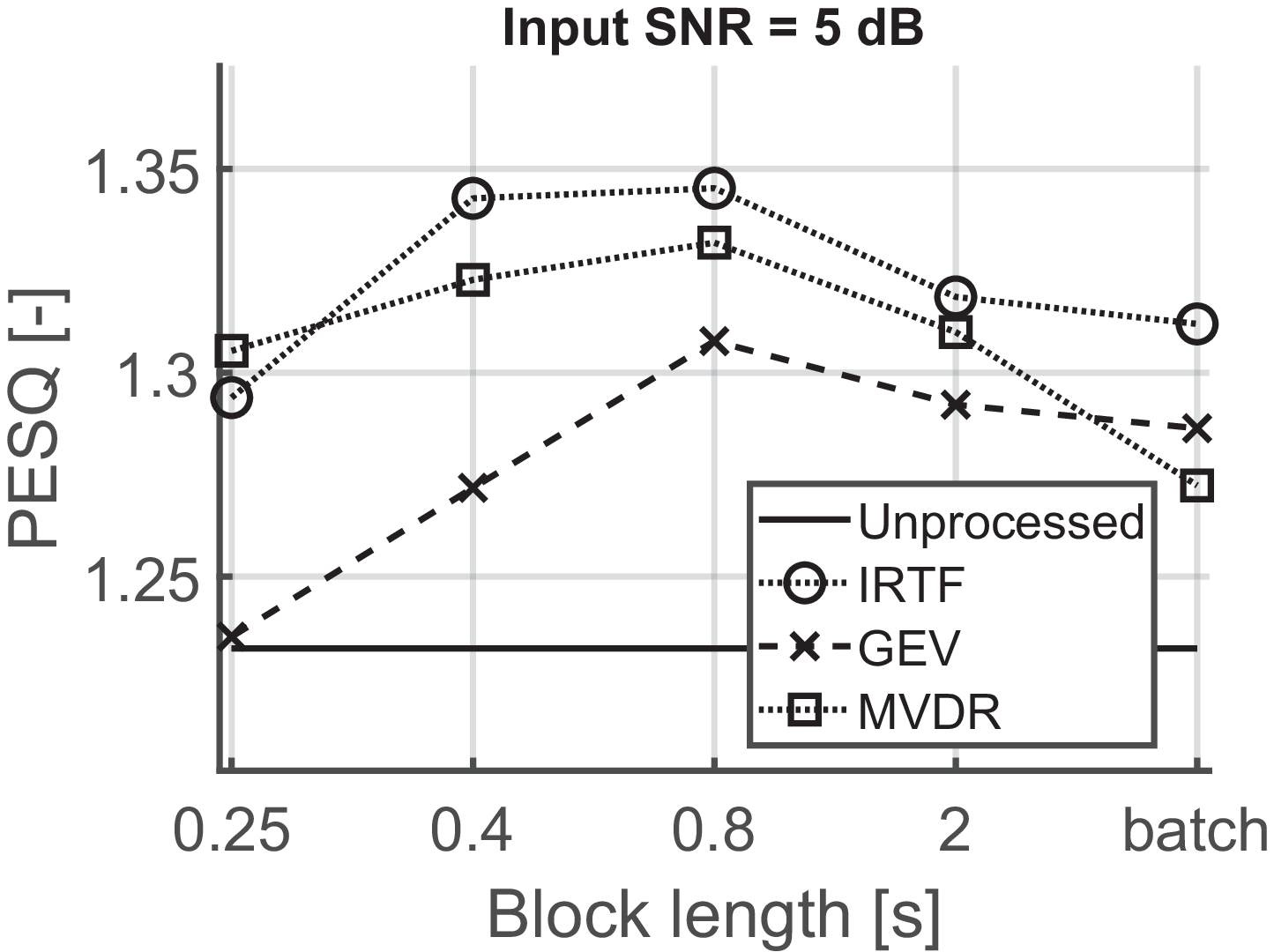}
	\includegraphics[width=0.49\linewidth]{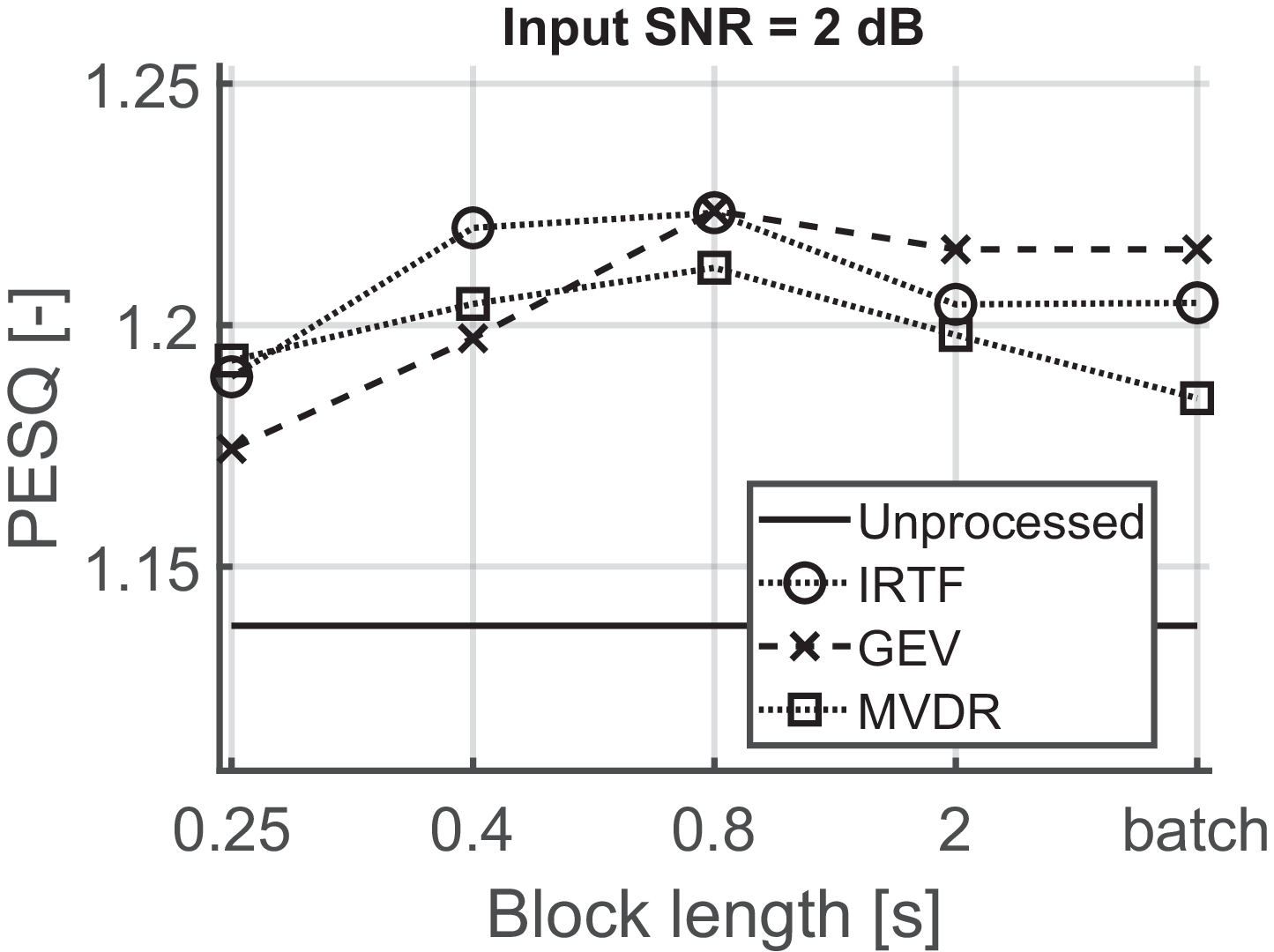}
	\includegraphics[width=0.49\linewidth]{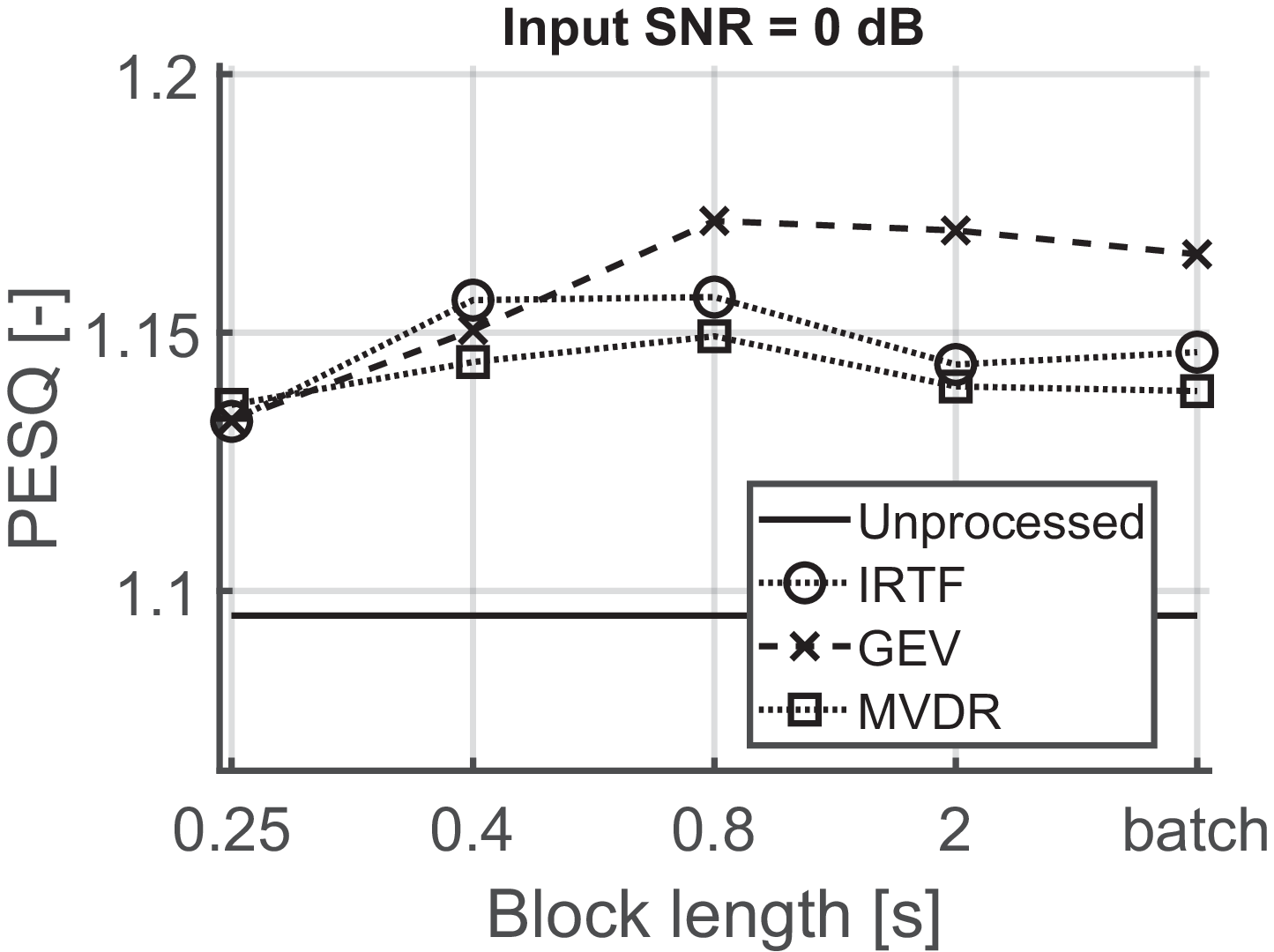}
	\caption{ \label{fig:pesq_lensnr_dyn} Dynamic: Perceptual quality measured by PESQ as function of the input SNR and length of the processed block. The proposed beamformers use mVAD detector and mask pooling.}
	\end{center}	
\end{figure}

The influence of input SNR on the enhancement is studied in Figure~\ref{fig:pesq_lensnr_dyn} using PESQ, which evaluates both noise suppression and speech distortion in a single measure.
The overall character of the results does not change. The proposed method is generally superior using the short blocks. The best PESQ is achieved for the block length $0.8$~s and it deteriorates for longer blocks due to the movements of the speaker. For input $\text{SNR}=0$~dB, noise attenuation is more crucial to the perceptual quality than low distortion. Here, GEV achieves the highest PESQ, which points to its strong noise suppression ability. With increasing input SNR, the proposed method yields higher PESQ in most cases, because it does not distort the target speech much.

\subsection{Influence of post-filtering}

\begin{figure}
    \begin{center}
	\includegraphics[width=0.49\linewidth]{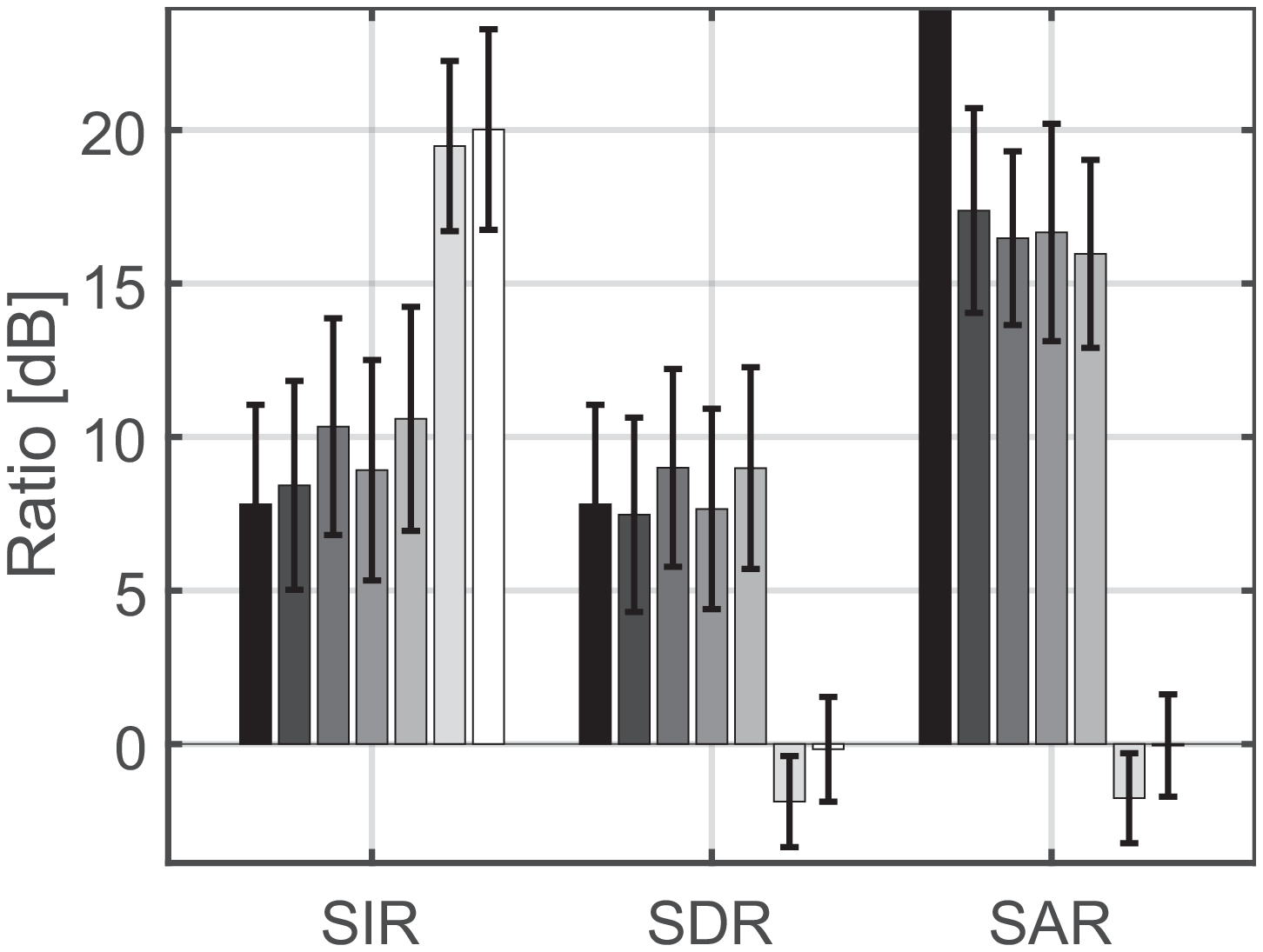}
	\includegraphics[width=0.49\linewidth]{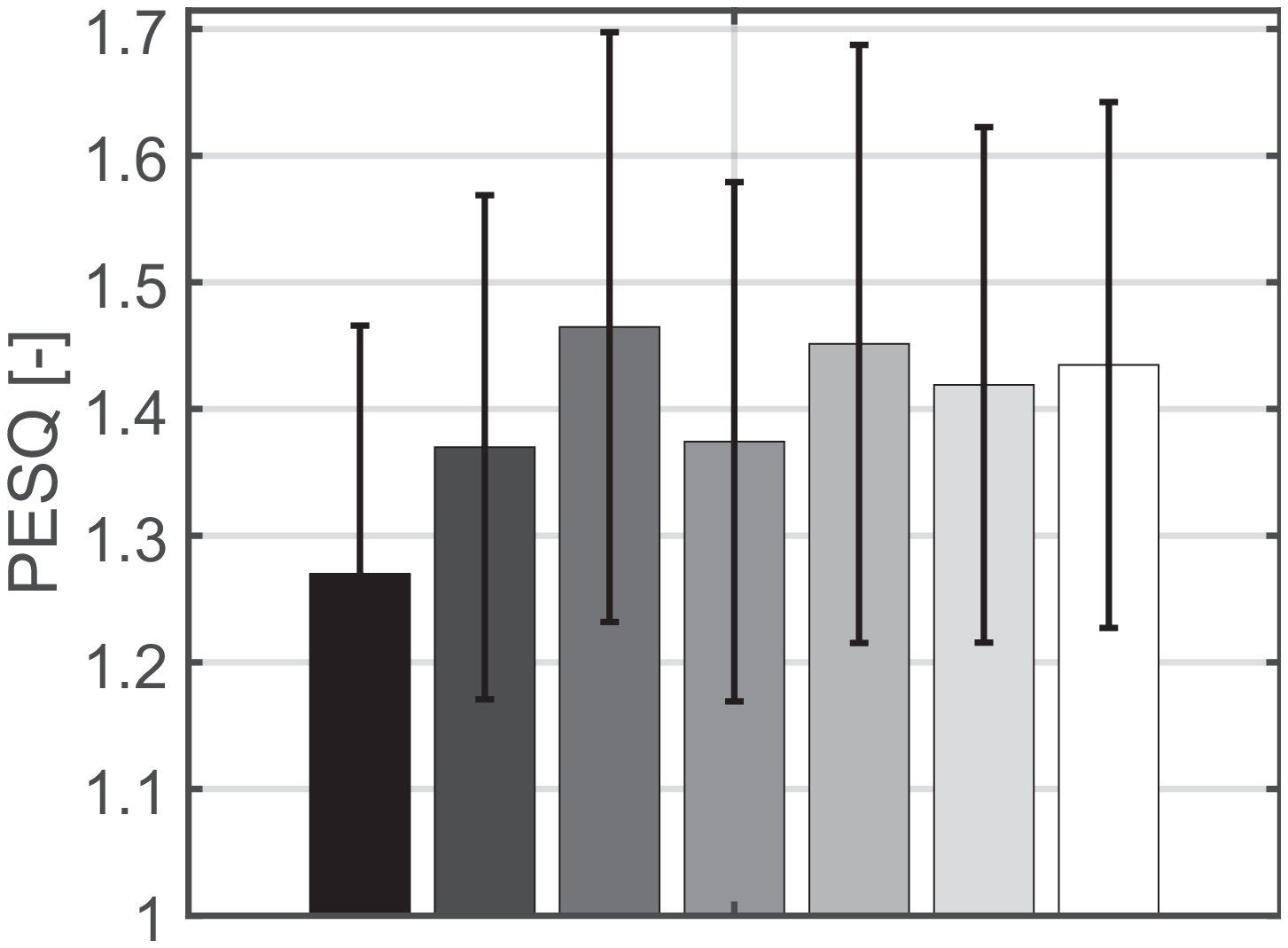}
	\includegraphics[width=0.80\linewidth]{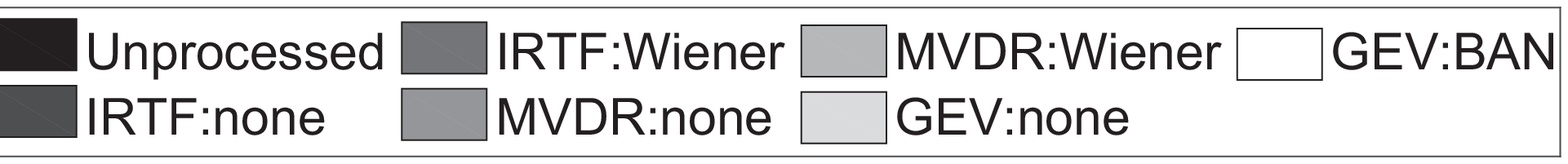}
	\caption{ \label{fig:bss_post} CHiME-4: Influence of post-filtering in the terms of 
		mean objective criteria and PESQ along with respective standard deviations. The proposed beamformers use mVAD detector and mask pooling. The SAR 
		values for unprocessed data are theoretically infinity, thus are 
		truncated in the graph.
	}
	\end{center}
\end{figure}

\begin{figure}
    \begin{center}
	\includegraphics[width=0.49\linewidth]{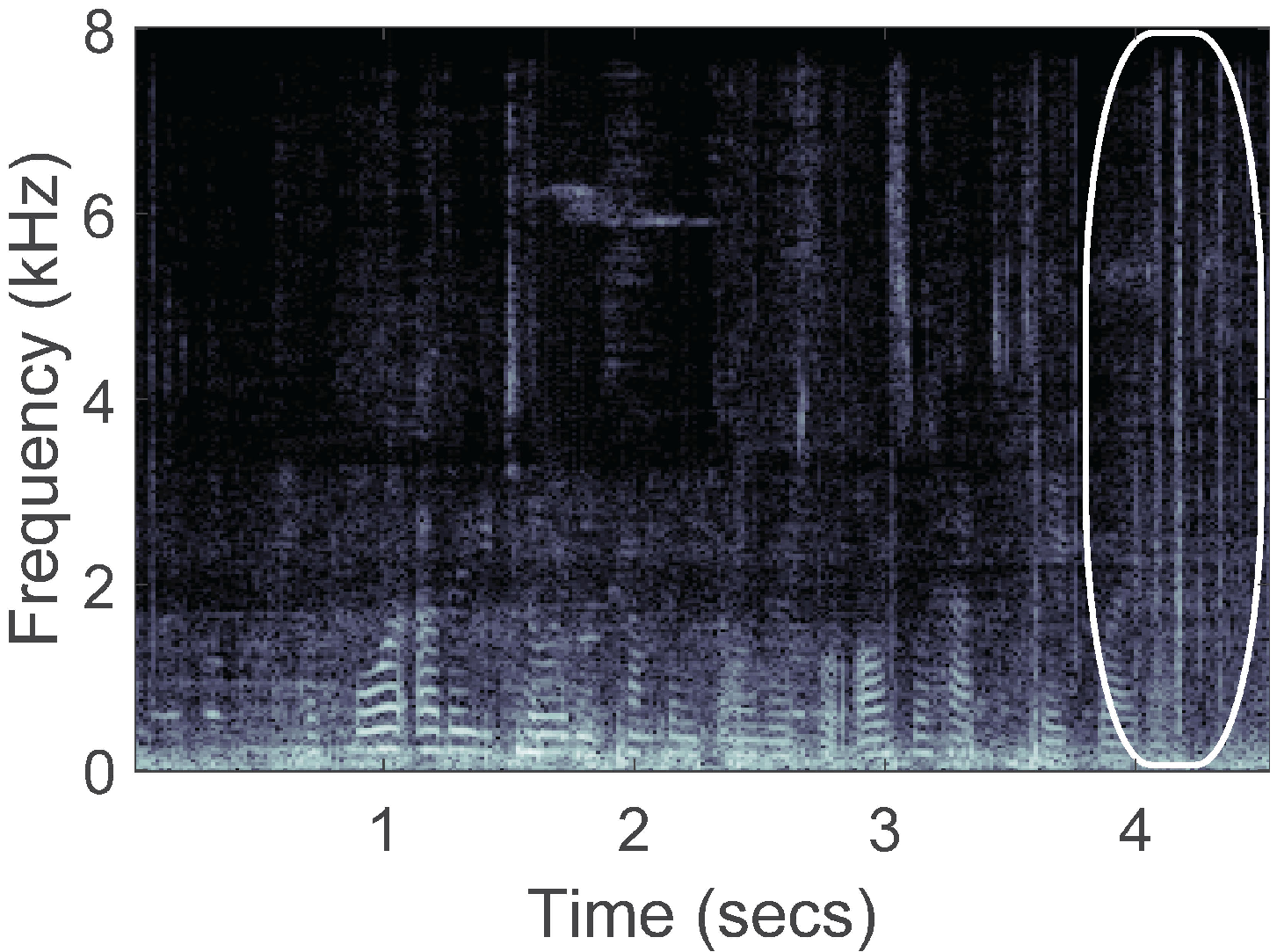}
	\includegraphics[width=0.49\linewidth]{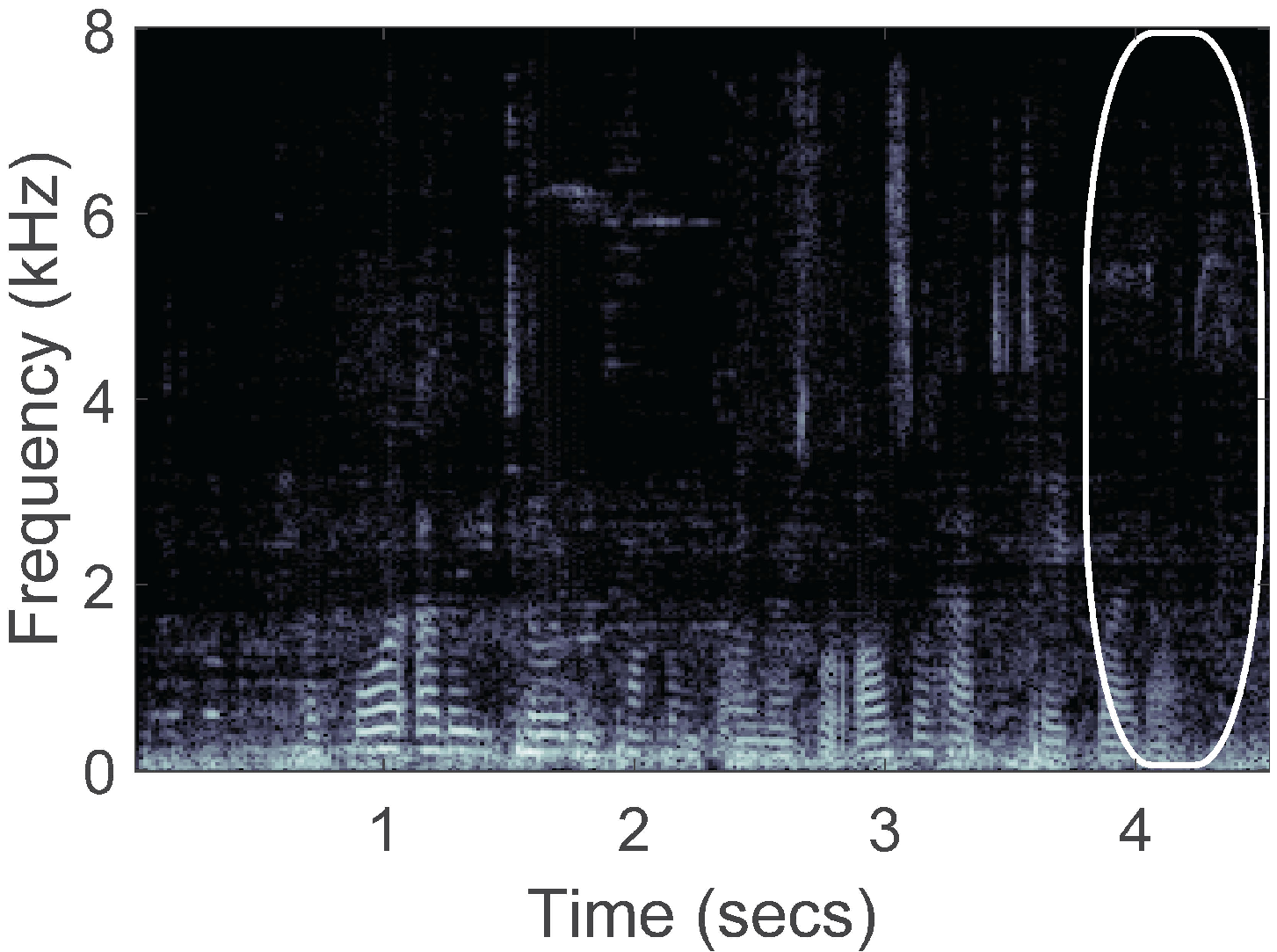}	
	\caption{\label{fig:spec_post} Demonstration of post-filter effects. On the left, there is signal without post-filtering, on the right is signal with applied post-filter ($f_{\rm max}$ was selected $8$~kHz to emphasise the effects of the post-filter). The residual noise and the highlighted artefact around time $4$~s are suppressed.}
	\end{center}
\end{figure}

In this experiment, the compared methods are tested with or without their 
post-filtering parts. The block-online regime with a block 
length of $0.8$s is considered together with mVAD and pooling of VAD masks.
The results achieved with this setting on CHiME-4 datasets are shown in~Table~\ref{tab:wer_post} 
and in Fig.~\ref{fig:bss_post}; an example of a spectrogram demonstrating the effects of post-filtering is shown in Figure~\ref{fig:spec_post}. The overall results show that the post-filtering improves both 
the perceptual quality as well as the accuracy of recognition.
When post-filtering is omitted, the WER is increased by about $0.5$\% for 
IRTF as well as for MVDR. The influence of BAN on the performance of GEV is more 
significant: Without BAN, the WER of GEV is increased by $1.7-3.8$\%.

For MVDR and IRTF, the omission of the Wiener post-filtering results in a decrease of SIR by about $2$dB and of SDR by about $1$dB. 
Although the SAR is slightly increased ($0.8$dB), the PESQ value is decreased by about $0.1$. 
The performance of GEV:BAN improves in the terms of SIR, SDR as well as SAR, by about $2$dB compared to GEV:none, 
and PESQ is improved by $0.03$. 

\begin{table}
    \begin{minipage}{0.47\textwidth}
	\caption{ \label{tab:wer_post} CHiME-4: Influence of post-filtering in terms of the WER (\%). The  proposed
		beamformers use mVAD detector and mask pooling.
		The best achieved  results are written in bold.}	\begin{center}
	\begin{tabular}{|c|c|c|c|c|}
		\hline
		\multirow{2}{*}{\begin{tabular}[c]{@{}c@{}}System:\\ post-filter 
		\end{tabular}} & \multicolumn{2}{c|}{Dev} &
		\multicolumn{2}{c|}{Test}\\
		& real & simu & real & simu\\
		\hline
		Unprocessed  & 9.83 & 8.86 & 19.90 & 10.79\\
		IRTF:none & 7.02 & \textbf{6.98} & 13.34 & 7.36\\
		IRTF:Wiener & \textbf{6.76} & \textbf{6.98} & \textbf{12.61} & 
		\textbf{6.95}\\
		MVDR:none & 7.03 & 7.95 & 14.72 & 7.97\\
		MVDR:Wiener & 6.85 & 7.50 & 14.05 & 7.47\\
		GEV:none & 10.02 & 9.49 & 19.52 & 10.27\\
		GEV:BAN & 7.87 & 7.78 & 15.71 & 7.94\\		
		\hline
	\end{tabular}
	\end{center}
	\end{minipage}
\end{table}

\subsection{Influence of pooling and VAD inclusion}
\label{sec:exp:vad}

This subsection investigates the effects of VAD output pooling on beamforming. Moreover, the contribution of VAD itself to the enhancement performance is measured. To this end, the results of beamforming with included/disabled VAD-mask weighting are compared, as discussed in Section~\ref{sec:rtf} below \eqref{eq:PSDestim}.

The results were achieved on CHiME-4 datasets in block-online regime with the block length $0.8$s without post-filtering. 

The results shown in Table~\ref{tab:wer_pool} suggest that the 
pooling is beneficial for the MVDR beamformer. Here, the omission of pooling 
increases the WER by $0.8-1.5$\%. Pooling does not improve (but neither deteriorate) the WER of IRTF. 
The objective measures did not show any significant benefits of pooling; that is why they are not presented.

Table~\ref{tab:wer_vad} indicates that the utilisation of mVAD is beneficial for the beamforming.
IRTF with mVAD performs better by $0.1-0.4$\% 
compared to IRTF:none, while MVDR is improved by $0.4-0.9$\% by utilisation of VAD.
Again, the objective criteria are not presented, since these do not reflect the inclusion of VAD significantly.

\begin{table}
    \begin{minipage}{0.47\textwidth}
    \begin{center}
    \caption{ \label{tab:wer_pool} CHiME-4: Influence of VAD mask pooling in the terms of WER (\%). 
    The beamformers use mVAD detector and do not use post-filtering. The best achieved results are written in bold.}
	\begin{tabular}{|c|c|c|c|c|}
		\hline
		\multirow{2}{*}{\begin{tabular}[c]{@{}c@{}}System:\\ pooling 
		\end{tabular}} & \multicolumn{2}{c|}{Dev} & 
		\multicolumn{2}{c|}{Test}\\
		& real & simu & real & simu\\
		\hline
		Unprocessed  & 9.83 & 8.86 & 19.90 & 10.79\\
		IRTF:no & 7.08 & 7.02 & \textbf{13.16} & \textbf{7.32}\\
		IRTF:yes & \textbf{7.02} & \textbf{6.98} & 13.34 & 7.36\\
		MVDR:no & 7.82 & 8.86 & 16.16 & 9.06\\
		MVDR:yes & 7.03 & 7.95 & 14.72 & 7.97\\
		\hline
	\end{tabular}
    \end{center}
    \end{minipage}
\end{table}

\begin{table}
    \begin{minipage}{0.47\textwidth}
    \begin{center}
	\caption{ \label{tab:wer_vad} CHiME-4: Influence of the VAD on beamforming in terms of the WER (\%).
		The techniques use neither pooling of VAD masks nor post-filtering.
		The best achieved results are written in bold.}	    
	\begin{tabular}{|c|c|c|c|c|}
		\hline
		\multirow{2}{*}{\begin{tabular}[c]{@{}c@{}}System:\\ VAD type 
		\end{tabular}} & \multicolumn{2}{c|}{Dev} & 
		\multicolumn{2}{c|}{Test}\\
		& real & simu & real & simu\\
		\hline
		Unprocessed  & 9.83 & 8.86 & 19.90 & 10.79\\
		IRTF:none & 7.21 & 7.34 & 13.56 & 7.52\\
		IRTF:mVAD & \textbf{7.08} & \textbf{7.02} & \textbf{13.16} & \textbf{7.32}\\
		MVDR:none & 8.19 & 9.26 & 17.06 & 9.67\\
		MVDR:mVAD & 7.82 & 8.86 & 16.16 & 9.06\\
		\hline
	\end{tabular}
    \end{center}
    \end{minipage}
\end{table}

\subsection{Computational aspects}
\label{sec:compreq}

To evaluate the computational burden due to the compared methods, we measure 
the time necessary to enhance the simulated development dataset in the bus 
environment. The experiment was performed on a PC with Intel processor 
\mbox{i$7$-$2600$K}, $3.4$GHz, and $16$GB RAM.
To maintain comparable conditions, only techniques implemented in 
Matlab are compared. Moreover, only a single computational thread is enabled during the experiments. All competing techniques are tested 
with mVAD (including GEV) or without any VAD 
(excluding GEV, which requires VAD to work properly).

\begin{figure}
    \begin{center}
	\includegraphics[width=0.49\linewidth]{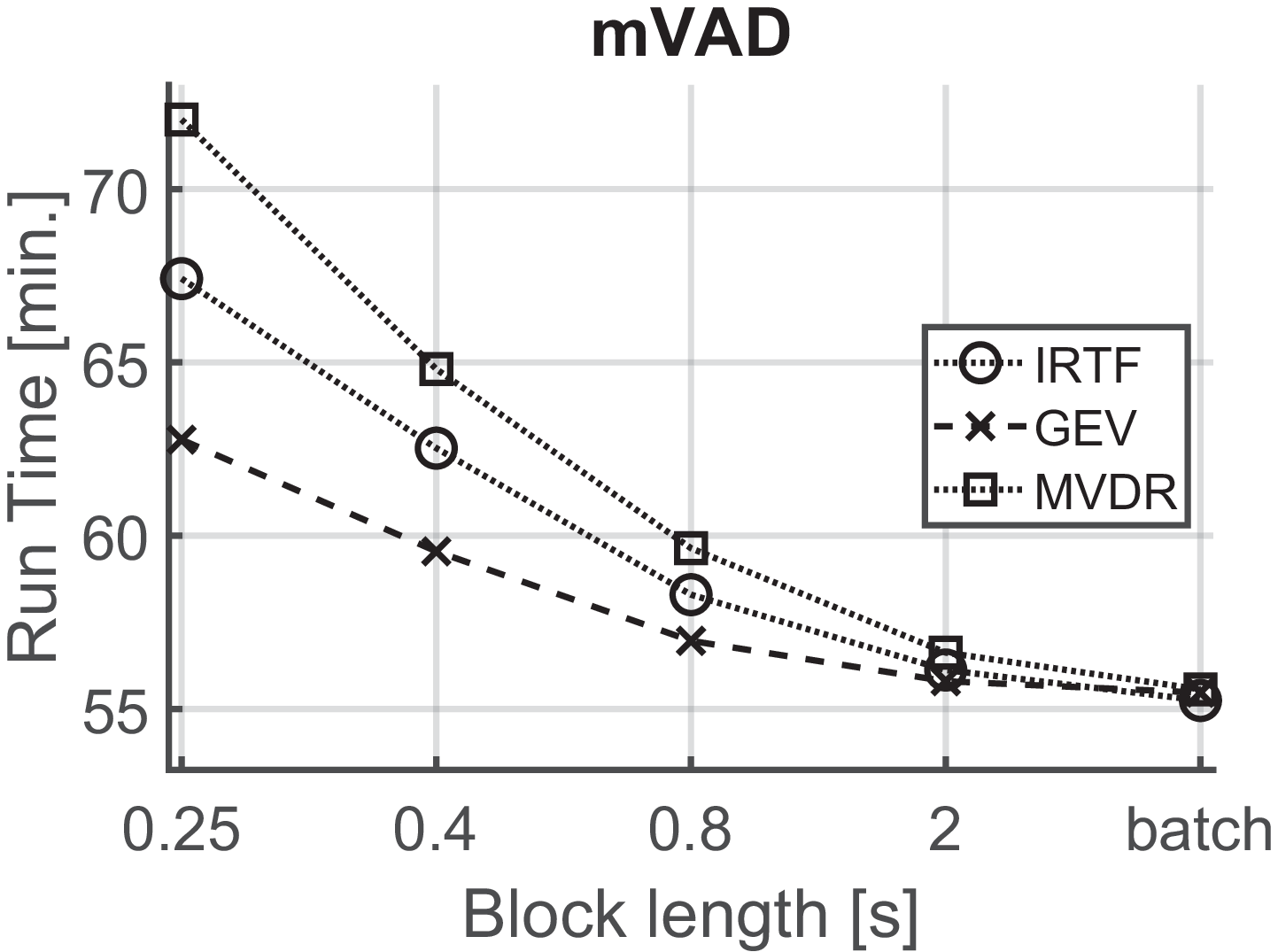}
	\includegraphics[width=0.49\linewidth]{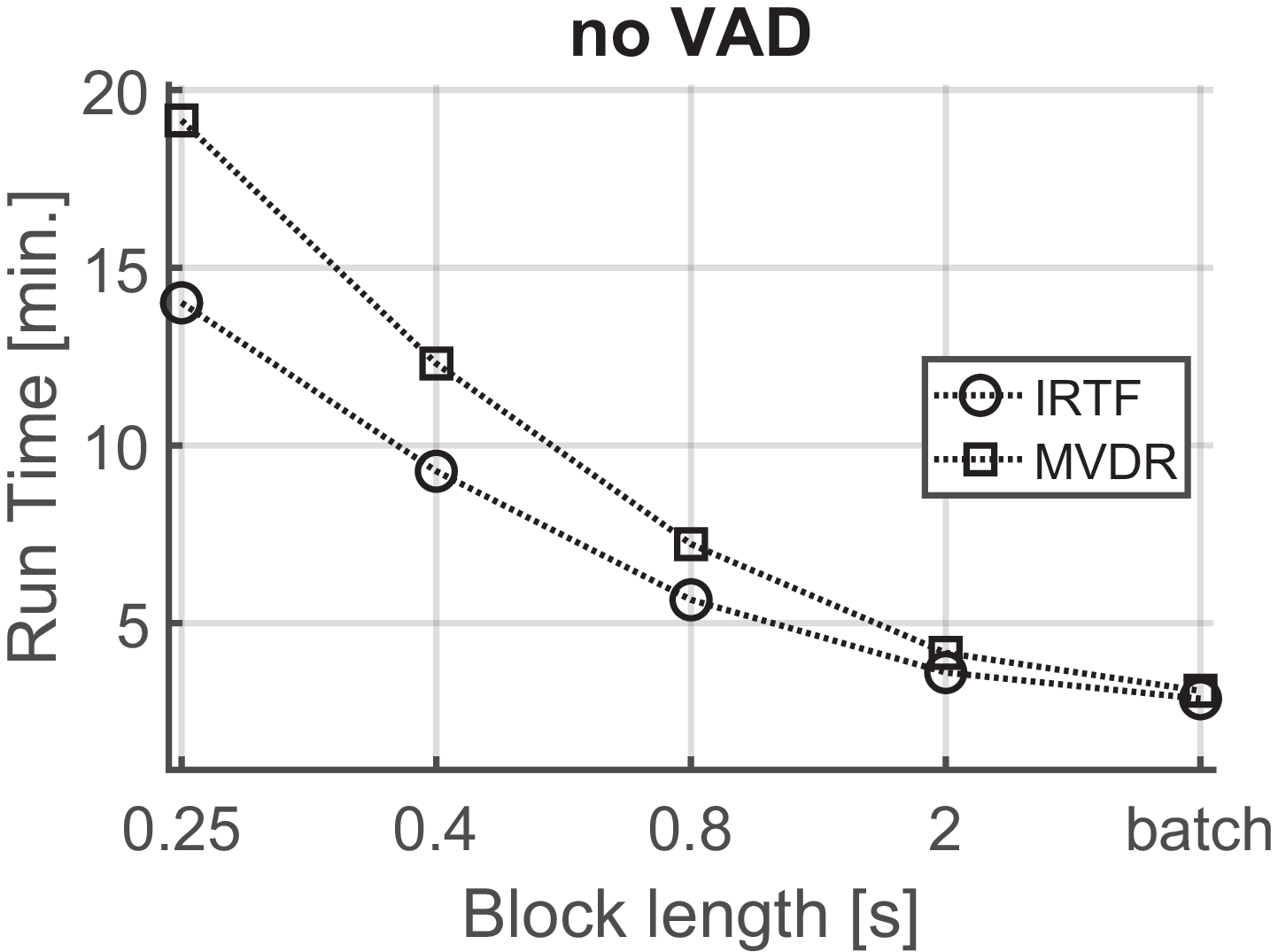}	
	\caption{\label{fig:bss_time} CHiME-4: Time required to enhance the 
		simulated development ``bus" dataset; the total length of the 
		recordings is $43.2$ minutes.}
	\end{center}
\end{figure}

Fig.~\ref{fig:bss_time} shows that the computational burden is 
decreasing with the growing length of block. This is, obviously, caused by 
the savings as the same processing is applied to longer segments of signals.
Most of the burden is due to VAD. In the case of batch processing, mVAD mask computation consumes about $90\%$ of the processing time for all techniques.

The results in Section~\ref{sec:exp:vad} indicate that the utilisation of VAD improves the performance of the presented technique. However, the method can be utilised without VAD if moderate performance loss is allowable. Thus, if the computational burden is of concern, the computational cost can be significantly reduced by omission of mVAD (or utilisation of a network with lesser number of parameters). 

\section{Conclusion}
\label{sec:conclusion}

A block-online multi-channel speech enhancement method based on beamforming has been proposed. 
In this method, the steering vector is constructed using estimates of RTFs between microphones. Performed experiments indicate that the performance of the proposed method is robust with respect to short length of the processed data. 

An alternative approach to computation of steering vector consists in eigenvalue decomposition of the speech covariance matrix.
In our experiments, the covariance-based techniques are represented by the state-of-the-art GEV beamformer. Considering the CHiME-4 datasets, the proposed methods achieve lower WER and higher PESQ values when blocks of length lower then $0.8$~s are used, whereas the GEV beamforming is superior for blocks of length $2$~s and higher. The proposed methods introduce significantly smaller amount of distortions as shown through the achieved SDR and SAR. 

The proposed IRTF beamformer appears to be computationally simpler than its MVDR counterpart, and it achieves lower WER and higher PESQ. We find beneficial to perform single channel post-filtering after beamforming; it improves the perceptual quality of the enhanced speech as well as it lowers the WER of the recogniser.

Concerning the future work, the proposed system (even without VAD) can serve as a strong initialisation for BSS methods, which can refine the resulting estimates of clean speech. 

\acuseall
\printacronyms[include-classes=abbrev,name=List of abbreviations,sort=true]

\section*{Acknowledgements}

This work was supported by The Czech Science Foundation through Project No. 17-00902S and by California
Community Foundation through Project No. DA-15-114599.

\bibliographystyle{iet}

\bibliography{mybib}

\end{document}